\pdfoutput=1

\documentclass[11pt]{article}

\usepackage{ACL2023}

\usepackage{url}
\usepackage{graphicx}
\usepackage{times}
\usepackage{latexsym}
\usepackage{amsmath}
\usepackage{hyperref}
\usepackage{xcolor}
\usepackage{array}
\usepackage{hhline}
\usepackage{colortbl}
\usepackage{latexsym}
\usepackage[capitalize]{cleveref}
\usepackage{microtype}
\usepackage{tabularx}
\usepackage{booktabs}
\usepackage{multirow}
\usepackage{xspace}
\usepackage{floatrow}
\usepackage{enumitem}
\setlist[enumerate]{itemsep=0mm}

\usepackage{subfig}
\usepackage{worldflags}
\usepackage[T1]{fontenc}
\usepackage[utf8]{inputenc}

\usepackage{microtype}

\usepackage{inconsolata}

\title{Uplifting Lower-Income Data: Strategies for \\Socioeconomic Perspective Shifts in Large Multi-modal Models}

\author{Joan Nwatu  \hspace{5pt} Oana Ignat \hspace{5pt} Rada Mihalcea \\
         University of Michigan - Ann Arbor, USA \\ \textit{\{jnwatu, oignat, mihalcea\} @umich.edu} \\ }

\usepackage{titlesec}
\titlespacing*{\section}{0pt}{0.1\baselineskip}{0.1\baselineskip}
\titlespacing*{\subsection}{0pt}{0.1\baselineskip}{0.1\baselineskip}

\begin{document}
\maketitle
\begin{abstract}
Recent work has demonstrated that the unequal representation of cultures and socioeconomic groups in training data leads to biased Large Multi-modal(LMM) models. To improve LMM model performance on underrepresented data, we propose and evaluate several prompting strategies using non-English, geographic, and socioeconomic attributes. We show that these geographic and socioeconomic integrated prompts favor retrieving topic appearances commonly found in data from low-income households across different countries leading to improved LMM model performance on lower-income data. Our analyses identify and highlight contexts where these strategies yield the most improvements. 
\end{abstract}

\section{Introduction}


A lack of diversity in popular AI datasets \cite{Shankar2017NoCW} leads to unequal model performance, further widening the technological gap between well-represented and underrepresented communities. While data from higher-income Western communities are readily available online, lower-income and non-Western data are often missing \cite{roslingfactfulness}. As a result, cost-effective methods like web scraping fail to produce diverse datasets.


One approach to building large datasets leverages LMM models to filter uncurated data based on image-text association strength scores \cite{fang2023data}. For instance, OpenAI's ViT-B/32 ~\cite{Radford2021LearningTV} was used to filter web-scraped images to create the LAION-5B dataset\cite{Laion2022}. However, foundation LMM models like CLIP perform unequally across cultures and socioeconomic groups, favoring higher-income and Western images \cite{nwatu-etal-2023-bridging}.



 \begin{figure}[h]
    \centering
    \includegraphics[width=0.9\textwidth]{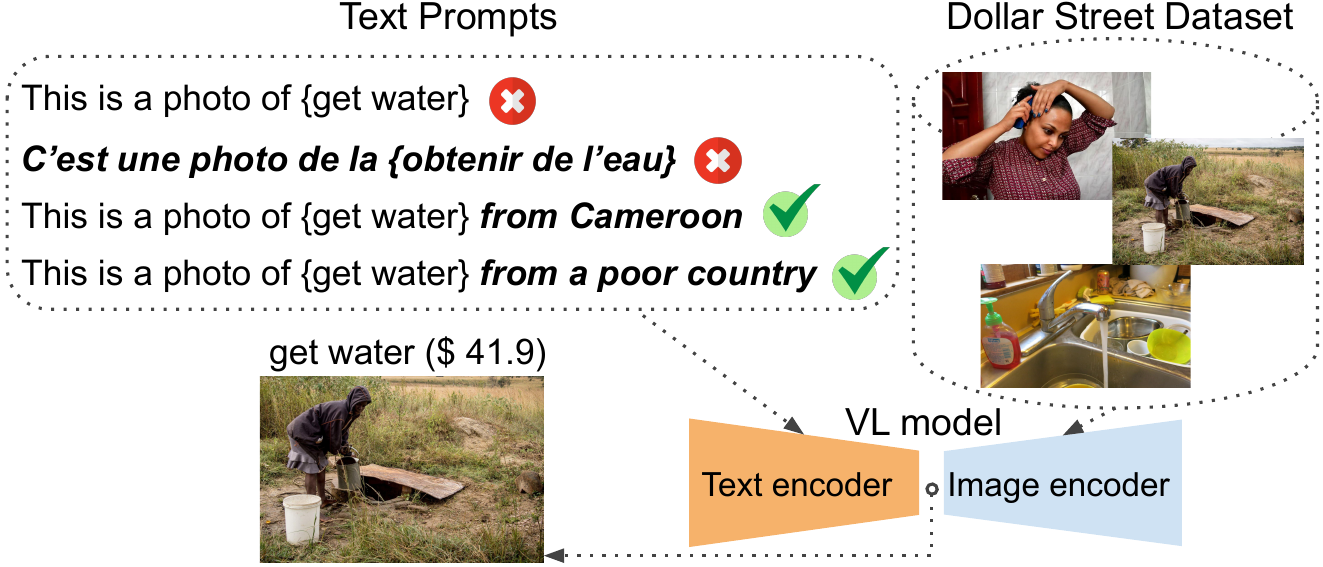}
    \caption{Low-income Image Retrieval from Dollar Street dataset~\cite{Rojas2022TheDS} using different prompt formulations. Prompts with integrated \textit{country} and \textit{income} information successfully retrieve fewer standard images previously left out by the English and translated (French) prompts.}
    \vskip -0.1in
    \label{fig:main}
\end{figure}

Datasets filtered by LMM models reflect the model's biases \cite{fang2023data}, often excluding underrepresented data and worsening the lack of diversity in AI models. \citet{ignat2024annotations} demonstrates this by showing that the LAION-5B dataset closely resembles data from Western countries, such as the United States and Canada while differing from non-Western countries' data. 
This leads to LMM models with uneven performance on data drawn from different locations and income groups.
Therefore, our paper seeks to answer the following question: \textit{\textbf{How do we improve the performance of LMM models on lower-income and non-Western data?}}


We tackle performance inequality in LMM models~\cite{Radford2021LearningTV, visheratin2023nllb} through prompting that transfers the cultural knowledge embedded in language \cite{ventura2023navigating, buettner2024incorporating, nguyen2024multilingual}. Our goal is to improve the performance of LMM models on data from households with non-Western and lower socioeconomic status.
Specifically,  as shown in \cref{fig:main}, we pose several research questions to evaluate the role of non-English languages, as well as prompts with geographic and socioeconomic attributes, to retrieve more diverse images.


Our contributions are summarized as follows. 
First, we show that a naive \textbf{prompt translation-based approach fails} to adequately address the performance gap of LMM models on lower-income data. Second, we establish that \textbf{geographic and socioeconomic attribute integrated prompts improve} LMM performance on lower-income data. We identify contexts where these prompts work best by conducting an in-depth analysis of LMM models' understanding of these attributes and their effects on recall across data from different countries. Lastly, we share insights from our analysis demonstrating how \textbf{these attributes drive a perspective shift} that benefits the retrieval of lower-income data.

\section{Related Work}

\paragraph{Addressing AI Performance Inequality.}

Class imbalances in training data contribute significantly to bias in AI models \cite{ferrara2024butterfly, Shankar2017NoCW, he2009learning, pouget2024no}, leading to unequal outcomes in areas like facial recognition \cite{ Buolamwini2018GenderSI}, healthcare \cite{obermeyer2019dissecting}, and hiring \cite{raghavan2020mitigating}. Since creating balanced datasets is challenging and costly \cite{ignat2024annotations, Ramaswamy2023BeyondWC}, researchers have explored bias mitigation techniques such as data augmentation, feature importance tuning, regularization, and adversarial training \cite{yan2020fair, zafar2017fairness, ignat2024annotations, maudslay2019s, Sharma_dataaug,navarro2024data, zhang2018mitigating}.
Our work is most similar to research on post-processing methods \citet{ferrara2023fairness, hardt2016equality, Kamiran_discriminationAware, pleiss2017fairness} that adjust model outcomes to meet diversity standards, aiming to benefit disadvantaged groups. Prior research has shown that LMM models perform poorly on data from lower socioeconomic groups, and our analysis investigates non-invasive post-processing methods to address this issue.

\paragraph{Multilingual AI Models.}

Language plays a key role in transmitting cultural knowledge \cite{callies2024cultural, sharifian2014language, karsdorp2019cultural, norton1997language}, as AI models often absorb biases from the language in their training data \cite{stanczak2021survey, rogers2021primer} and model outputs can be controlled by specifying a cultural shift in perspective \cite{ventura2023navigating} to improve diversity.
However, research \citet{arora-etal-2023-probing,cao-etal-2023-assessing, alkhamissi2024investigating, liu2021visually} shows that large language models (LLMs) and LMM models capture more cultural information from English data (mainly Western) than from non-English data. This disparity stems from differences in the quantity and quality of non-English data, translation issues, and model design \cite{arora-etal-2023-probing, hershcovich-etal-2022-challenges, nasif1991methodological}.

Similar to past studies \cite{Devries2019DoesOR, nguyen2024multilingual} using multilingual approaches to enhance data diversity, our work explores how multilingual large multi-modal models and non-English languages can improve representation across regions and income groups.

\paragraph{Prompting AI Models.}

Recent studies have explored prompting techniques for large language models, including both hard \cite{petroni-etal-2019-language, zhou2023controlled} and soft prompting \cite{huang2023diversity, goswami2023contextual}, to improve model adaptation for tasks like instruction tuning, and value alignment. These methods are also applied in LMM models \cite{lu2022prompt, yao2024cpt, zhou2022learning}. 
While prior work \cite{buettner2024incorporating}  has incorporated geographic and physical attributes into prompts to enhance image retrieval diversity, this research extends the investigation to non-English language prompts and socioeconomic attributes to analyze how LMM models encode representations of various topics across regions and socioeconomic status.


 \section{Methodology}
We propose prompting strategies that account for language, location, and socio-economic attributes and analyze how these prompts affect the performance of a multilingual LMM model on data across different socio-economic groups, primarily focusing on lower-income data.

\subsection{Dollar Street Dataset} \label{dataset}
We use the Dollar Street~\cite{Rojas2022TheDS}, which contains $38,479$ images of household items (e.g., ``stoves'', ``cutlery'', ``toothbrush'') spanning a large number of countries and several income levels. The dataset images were sourced from households in $63$ countries on four continents (Africa, America, Asia, and Europe). The number of images ranges from $45$ in Canada to $4,704$ in India, with a median of $407$ images per country. Size and image resolutions vary slightly across data from different regions; however, the mean and median image properties per region are relatively similar.


\paragraph{Image Income Classes.} Each image is accompanied by the monthly household income value in U.S. dollars, calculated to reflect monthly consumption and adjusted for purchasing power parity to match the variance in cost of living across the different regions. The monthly income values range from $26.9\$$ to $19,671.0\$$.

For fair comparison across bins, we group the images using the quartile binning method, which splits the data into an approximately equal number of images per bin as shown in \citet{Rojas2022TheDS}. 
We group the images into four income classes (``poor'', ``low-mid'', ``up-mid'', and ``rich'' ) using quartiles as shown in \cref{tab:data_split_income}. We further categorize the lowest two image income classes as {\it lower-income images} and the highest two income groups as {\it higher-income images}. 



\vspace*{-0.1\baselineskip}

\begin{table}[h]
\centering
\resizebox{0.6\columnwidth}{!}{%
\begin{tabular}{c|c}
Quartile name& Income range \\
\midrule
poor    & 26.9 - 95.0          \\ 
low-mid & 195.4 - 685.0        \\ 
up-mid  & 694.0 - 1,998.0      \\ 
rich    & 2,001.0 - 19,671.0  
\end{tabular}%
}
\caption{Income quartiles and their ranges for all the images in Dollar Street.}
\label{tab:data_split_income}
\end{table}

\vspace*{-0.6cm}

\paragraph{Country Economic Classes.}
We group all 63 countries from Dollar Street into country economic classes based on their World Bank income classification.\footnote{\url{https://datahelpdesk.worldbank.org/}}
All the countries and their economic classes are shown in \Cref{tab:country_stats}. 
We further categorize the lowest two country economic classes as {\it lower-income countries} and the highest two economic groups as {\it higher-income countries}.

\paragraph{Topic Representations.} 
There are $291$ unique topics associated with the images in the dataset which reflect everyday household objects and human actions (e.g., ``toilet paper'', ``get water''), some of which are subjective (e.g., ``next big thing I plan to buy'', ``favorite sports clubs'', ``most loved item''). We remove nineteen subjective topics from the dataset following \citet{Devries2019DoesOR} and \citet{nwatu-etal-2023-bridging}.

\subsection{Prompt Design}
We describe below the prompting strategies we use for our experiments and show examples in Figure~\ref{fig:main}.

\paragraph{Default English Topic Prompt.}\label{default prompt}
Using the topics, we formulate an English prompt without any modifications (e.g., ``This is a photo of \textit{cutlery}''), as described in \citet{Radford2021LearningTV}, to which we refer to as the \textit{default English prompt}. The performance obtained using these prompts is set as our baseline.

\paragraph{Translated Topic Prompt.} \label{translated prompt}
For our multilingual experiments, we investigate the impact of non-English language prompts on the Dollar Street dataset. 
We use the term \textit{non-English major language}  to refer to the non-English language that is most widely spoken or most commonly used in a particular country or region.

Specifically, we pair each country with their non-English major language (e.g., \textit{Portuguese} for \textit{Brazil}, \textit{French} for \textit{Cameroon}) following the country and language information provided by official sources.\footnote{\url{www.cia.gov/the-world-factbook/field/languages/}, \url{www.ncsc.org/__data/assets/pdf_file/0024/17862/languagesbycountries.pdf}, \url{www.dss.gov.au/sites/default/files/files/foi_disclosure_log/12-12-13/language-list.pdf}} 

We identify 59/63 countries in Dollar Street where one or more major non-English languages are spoken. We also select languages covered by state-of-the-art machine translation and multilingual LMM models.
There are 40 such non-English major languages, and they are listed in \Cref{tab:country_stats}.

Finally, we translate the \textit{default English prompts} to these 40 languages using the NLLB-200-distilled-600M~\cite{costa2022no}, an open-source state-of-the-art neural machine translation model. Translation metrics for NLLB-200-distilled-600M are shown in Appendix \cref{tab:Translation metrics} and available on HuggingFace.
If an image prompt is translated into the non-English major language of the image's country of origin, it is referred to as a \textit{native translated prompt}.

\vspace*{-0.2cm} 

\paragraph{Country Suffix Topic Prompt.} \label{country_suffix}
For our second prompting technique, we include country names as suffixes to the default English prompt (e.g., ``This is a photo of \textit{cutlery} from \textit{Cameroon}''). We create 63 new prompt templates by adding the country names of each of the 63 countries in Dollar Street. 
We refer to these prompts as \textit{country-suffix prompts}.

\paragraph{Income Suffix Topic Prompt.} \label{income_suffix}
We also create prompts by integrating socio-economic attributes (e.g., ``poor country'', ``rich region'') as suffixes to the \textit{default English prompt}.  For instance,  a sample prompt is ``This is a photo of \textit{cutlery} from \textit{a rich country}''. 
For more robust results, we use multiple synonyms each for the {\it poor} and {\it rich} attributes (e.g., ``an impoverished country'', ``a wealthy region''). We also create prompts using neutral suffixes (e.g., ``a country'', ``a home'').  We refer to these prompts as \textit{income-suffix prompts}.

\subsection{State-of-the-art LMM Model}
For our evaluation, we chose NLLB-CLIP-SigLIP \cite{visheratin2023nllb}, a state-of-the-art multilingual LMM model, due to its broad reach across many low-resource languages and superior performance among other models.\footnote{\url{https://huggingface.co/visheratin/nllb-clip-large-siglip}}
The model consists of an image encoder from the SigLIP model \cite{big_vision, zhai2023sigmoid} and a text encoder from the NLLB model \cite{costa2022no}. The model supports the 201 languages of the Flores-200 \cite{costa2022no} and has recorded groundbreaking results on the Crossmodal-3600 dataset \cite{ThapliyalCrossmodal2022}, especially on low-resource languages.

\section{Research Questions}

We perform several analyses to answer three research questions that uncover and mitigate limitations in the performance of LMM models across different countries and socioeconomic groups. 

\subsection{RQ1. Do translated prompts improve retrieval performance for lower-income images?} \label{rq1}

We calculate the cosine similarities between image and translated prompt text embeddings for each image-topic pair across English and 40 non-English languages, generating 41 alignment scores per image. The alignment scores with \textit{default English prompts} serve as our baseline.

We compute Recall scores by selecting the top \textit{N} images with the highest alignment scores for each topic, where \textit{N} represents the number of ground truth images. We then group and analyze the Recall scores across different countries and image income classes and present our findings below.



\paragraph{\textit{Native translated prompts} perform\\ consistently worse than \textit{English prompts} on \\lower-income images from their respective countries.}

\begin{figure}[h]
    \centering
    \includegraphics[width=0.9\textwidth]{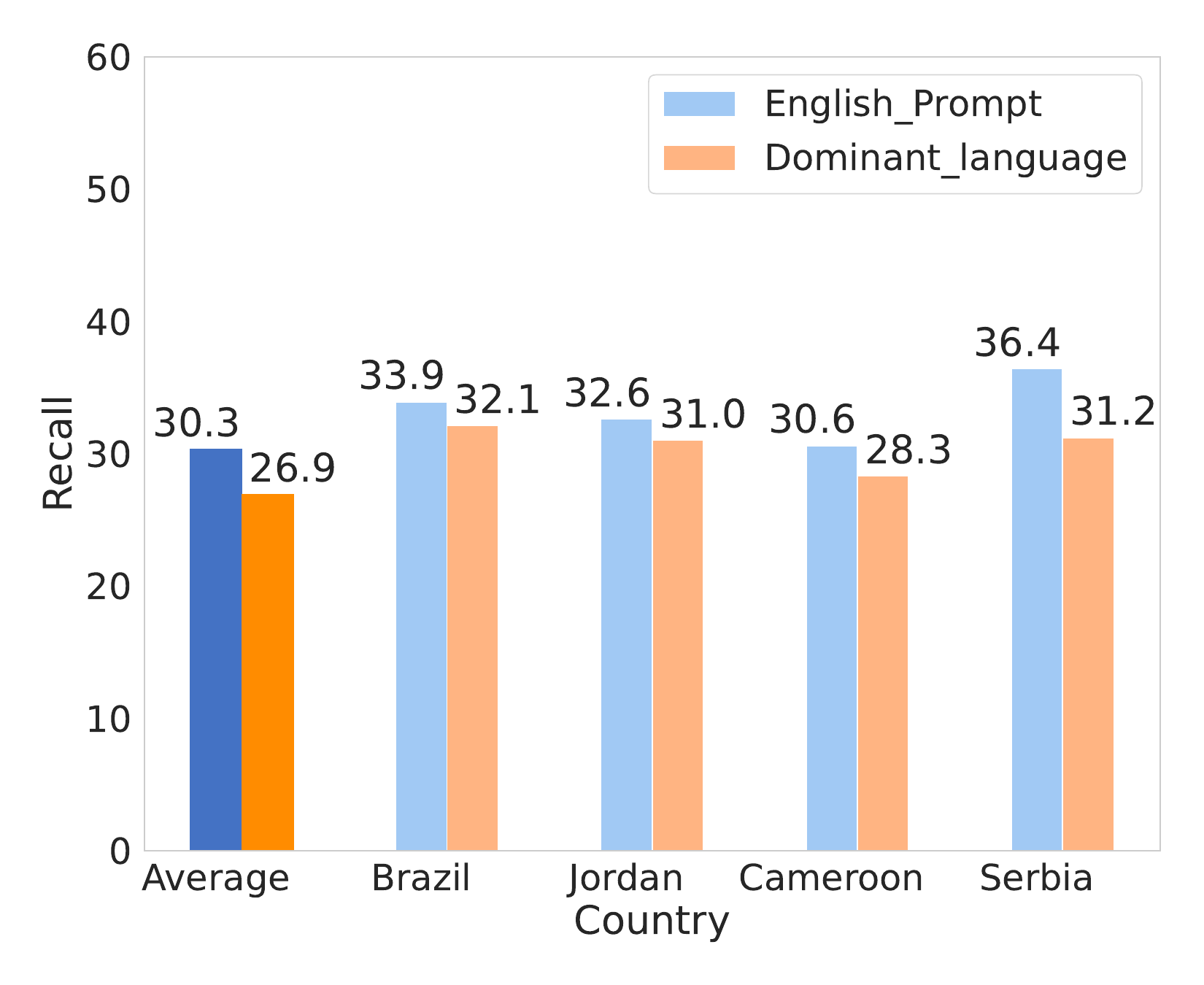}
    \caption{NLLB SigLIP Recall (\%) over poor and lower-middle income images from four countries, one from each of the four continents: Asia, Africa, America, and Europe for English and native translated prompts. \textit{Best viewed in color.}} 
    \vskip -0.1in
    \label{fig:Domisworse}
\end{figure}

We focus our analysis on images from the two lowest image income groups, i.e., \textit{poor} and \textit{low-middle} as grouped in \ref{dataset}. 
After excluding 20 countries without data for these income groups (e.g., \textit{Russia}, \textit{Turkey}), we retain 39/59 countries and 28/40 non-English languages for the study.
Each country is paired with its native non-English language, and we compare Recall scores for the \textit{native translated prompts} to those for the \textit{default English prompts}.
The average Recall across all countries and scores from four countries are displayed in \Cref{fig:Domisworse}.



For 35 out of 39 countries, the \textit{native translated prompts}  underperform compared to the \textit{default English prompts}. The exceptions include \textit{Burkina Faso}, \textit{Nigeria}, \textit{Pakistan} and \textit{Tanzania}, where \textit{native translated prompts} in \textit{French} (diff. of 1.0), \textit{Hausa} (diff. of 0.2), \textit{Urdu} (diff. of 0.7) and \textit{Swahili} (diff. of 1.5), respectively, outperform English prompts. Overall, native translated prompts generally fail to retrieve diverse images, as depicted by the example using French prompt in \cref{fig:main}.


\paragraph{The best-performing non-English language \\often differs from the country's native \\language.}

\begin{figure}[!ht]
\centering
{{\includegraphics[width=\textwidth]{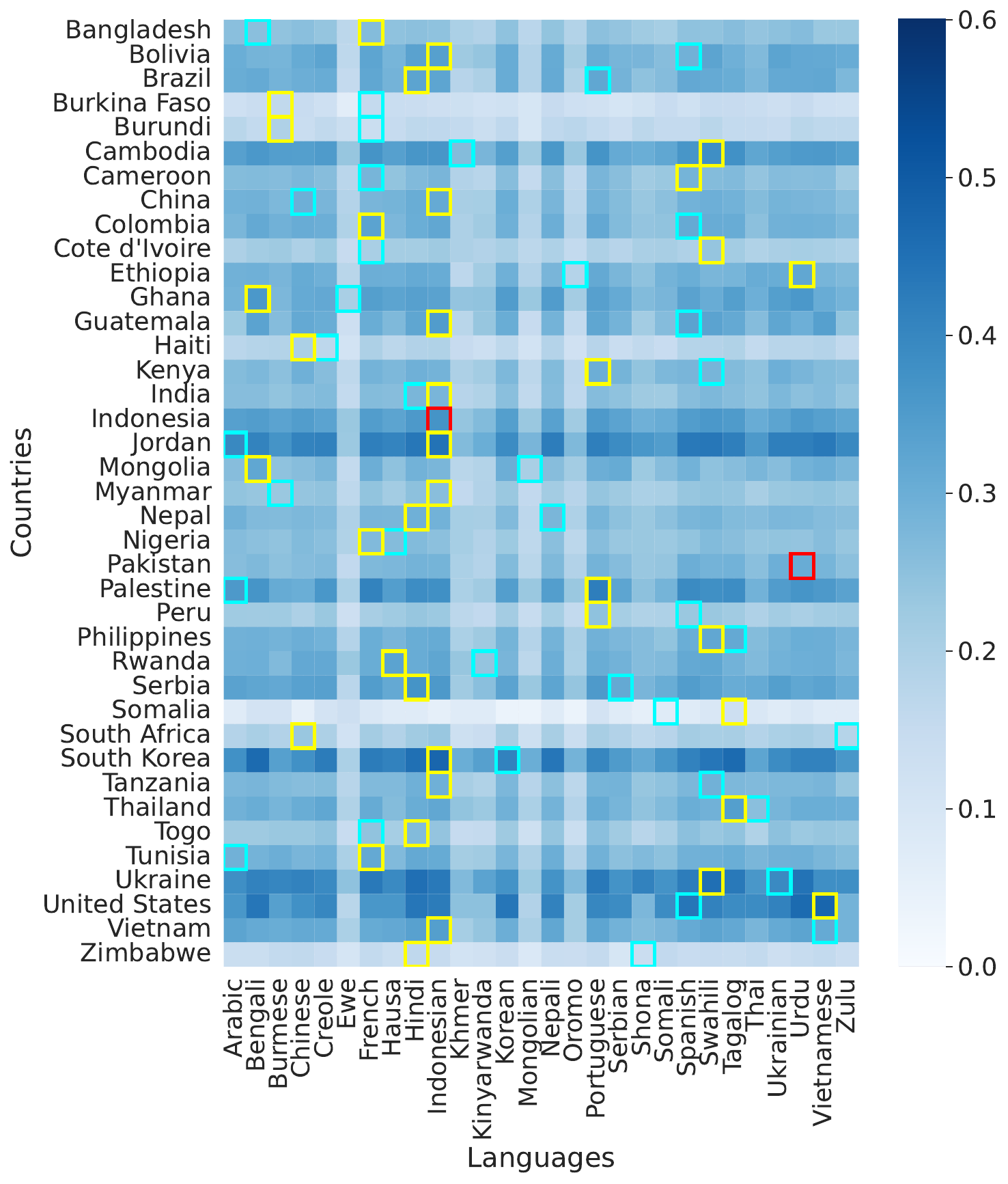} }}
    \caption{Recall scores for \textit{lower} income images from 39 countries and 28 languages. The cyan highlight shows the Recall for a country's native translated language, the yellow highlight shows the best-performing language recall, and the red shows the Recall for the language that is both the native and highest performing for that country. \textit{Best viewed in color}.  
}%
\vskip -0.1in
\label{fig:nllbheatmap}%
\end{figure}

We analyze the Recall scores for lower-income images across 28 language prompts used in different countries and find that the best-performing language prompts often differ from the countries' non-English major languages. Specifically, in 24 out of 39 countries, non-English language prompts outperform the default English prompts, yet these top-performing languages are not typically spoken in the respective countries. As illustrated in \Cref{fig:nllbheatmap}, for 37/39 countries, the language with the highest Recall score (highlighted in yellow) differs from the country's primary non-English language (highlighted in cyan), with exceptions in Indonesia and Pakistan, where they coincide (highlighted in bold red).


\paragraph{\textit{Translated prompts} decrease performance for\\ all image income classes across all\\ countries.}

We analyze the impact of 40 non-English language prompts on all images from Dollar Street, covering 59 countries and we group Recall scores by image income classes. By comparing Recall scores between  \textit{default English prompts} and \textit{native translated prompts}, we assess the effect of each non-English language on four income classes and show the difference in scores in Appendix \Cref{tab:drop_dom_lang_all}.


\begin{table}[h]
\centering
\resizebox{0.9\columnwidth}{!}{%
\begin{tabular}{c|cccc}

\multirow{3}{*}{\textit{\textbf{\begin{tabular}[c]{@{}l@{}}Non-English\\ languages\\ (Average)\end{tabular}}}} &
  \multicolumn{4}{c}{\textbf{Image Income Class}} \\ \cmidrule{2-5} 
 &
  \textbf{Poor $\Delta$} &
  \textbf{Low-mid $\Delta$} &
  \textbf{Up-mid $\Delta$} &
  \textbf{Rich $\Delta$} \\ 
 &
  20.2 \textcolor{purple}{(-2.2)} &
  31.0 \textcolor{purple}{(-4.9)} &
  37.8 \textcolor{purple}{(-7.8)} &
  36.1 \textcolor{purple}{(-7.5)} \\ 
\end{tabular}%
}
\caption{Average differences between Recall scores for non-English language prompts and Recall scores for default English prompts for all data, grouped by image income classes. We find that non-English prompts lead to a decrease in Recall scores across all income classes.}
\label{tab:drop_dom_lang}
\end{table}

We show in \cref{tab:drop_dom_lang} the average Recall and drops in performance across all 40 translated prompts for each image income class.
The results indicate that higher-income classes, specifically the \textit{rich} and \textit{up-mid} groups, experience the largest drops in performance with translated prompts. This may be due to the overrepresentation of images from these income groups in AI models and datasets, positioning them as the "standard" representation. Similarly, English, the dominant training language, is seen as the ``standard'' for textual data, so non-English prompts may signal a deviation from this standard, resulting in poorer model performance.


\subsection{RQ2. Does adding country information improve retrieval performance for lower-income images?} \label{rq2}

We compute cosine similarity scores between NLLB-CLIP-SigLIP image embeddings and the text embeddings of 63 \textit{country suffix prompts}., yielding 63 image-topic alignment scores per image. Using the alignment scores from the \textit{default English prompts} as a baseline, we follow the procedure outlined in Section \labelcref{rq1} to calculate Recall scores for each topic with the \textit{country suffix prompts}. 
We then analyze the impact of adding country suffixes to text prompts and present the results in the following sections.



\paragraph{\textit{Country-suffix prompts} perform consistently better than \textit{default English prompts} on lower-\\income images.}

\begin{figure}[ht]
    \centering
    \includegraphics[width=0.8\textwidth]{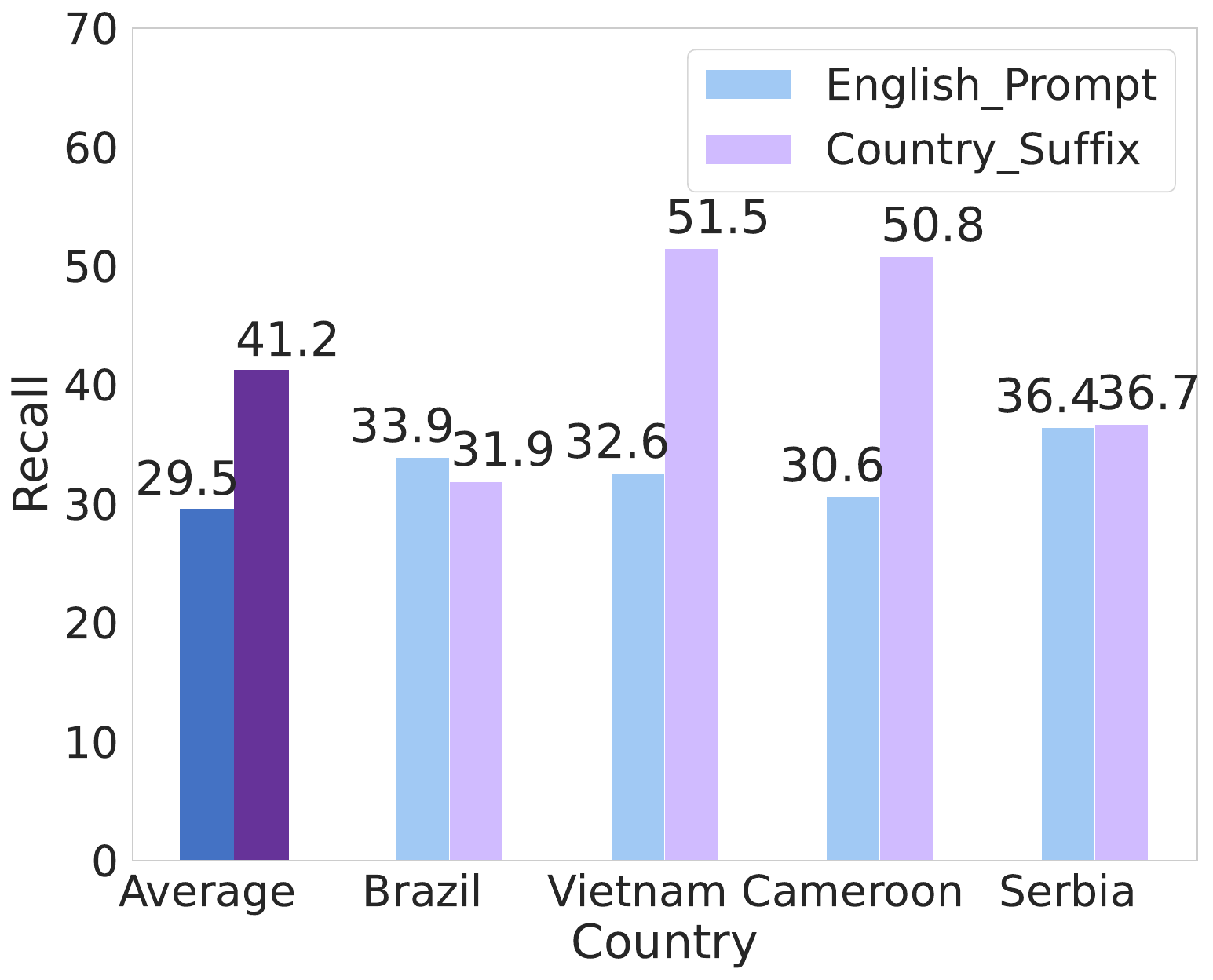}
    \caption{Recall (\%) with NLLB SigLIP over \textit{poor} and \textit{lower-middle} income images from four countries from Asia, Africa, America, and Europe, for English and Country Suffix prompts. \textit{Best viewed in color.}}
    \vskip -0.1in
    \label{fig:NLLBCountisbest}
\end{figure}

Focusing on low-income data, we filter out 21 countries without images from \textit{poor} or \textit{low-mid} income households, leaving 42 countries for analysis. In \Cref{fig:NLLBCountisbest}, we present the average Recall scores across all countries using both \textit{default English} and \textit{country-suffix prompts}, along with results from four sample countries from different continents.

Our findings indicate that in 38/42 countries, adding a \textit{country-suffix}  to text prompts improves Recall performance for lower-income images compared to \textit{default English prompts}. Exceptions include \textit{Bolivia, Brazil, Jordan}, and the \textit{United States}. \textit{Country-suffix prompts} are thus more effective in retrieving diverse images, as demonstrated by the Cameroon example in \cref{fig:main}.



\paragraph{A country's economic status influences the performance of its \textit{country-suffix prompt} across different image income classes.}

Country suffix prompts improve LMM model performance for lower-income images (\textit{poor}) but reduce performance for higher-income images. Using World Bank income classifications, we calculate Recall scores across four country suffix groups (poor, low-mid, up-mid, and rich) and four image income classes (based on household income). For each image income class, we aggregate Recall scores and compare them with those from default English prompts, as shown in \cref{tab:Effect-table}, with detailed results in Appendix \cref{tab:all_trend_table} and \cref{tab:trend_table_2}. For example,  \cref{tab:Effect-table} shows that Recall of images from poor households using \textit{country suffixes} of poor countries is 31.2, a 9.7 increase from \textit{default English prompt} performance on that group.

The analysis reveals that country suffixes from poor, low-mid, and up-mid income categories improve Recall for images from poor households, while reducing Recall for higher-income groups (low-mid, up-mid, rich).

\begin{table}[h]
\centering
\resizebox{0.85\textwidth}{!}{%
\begin{tabular}{c|cccc}

\multirow{2}{*}{\textbf{\begin{tabular}[c]{@{}l@{}}Country\\ suffix (Avg)\end{tabular}}} & \multicolumn{4}{c}{\textbf{Image Income Classes}} \\ \cmidrule{2-5}
 &
  \multicolumn{1}{l|}{\textbf{Poor} \hspace{2pt}  $\Delta$} &
  \multicolumn{1}{l|}{\textbf{Low-mid \hspace{2pt}  $\Delta$}} &
  \multicolumn{1}{l|}{\textbf{Up-mid \hspace{2pt}  $\Delta$}} &
  \textbf{Rich \hspace{2pt}  $\Delta$} \\ \midrule
Poor  &
  \multicolumn{1}{l|}{\textbf{31.2} \small{\textcolor{teal}{(+9.7)}}} &
  \multicolumn{1}{l|}{30.7 \small{\textcolor{purple}{(-5.3)}}} &
  \multicolumn{1}{l|}{25.5 \small{\textcolor{purple}{(-20.6)}}} &
  21.9 \small{\textcolor{purple}{(-22.5)}} \\ 
Low-mid   &
  \multicolumn{1}{l|}{29.2 \small{\textcolor{teal}{(+4.2)}}} &
  \multicolumn{1}{l|}{\textbf{31.6} \small{\textcolor{purple}{(-4.3)}}} &
      \multicolumn{1}{l|}{27.6 \small{\textcolor{purple}{(-15.3)}}} &
  23.2 \small{\textcolor{purple}{(-16.9)}} \\ 
Up-mid  &
  \multicolumn{1}{l|}{24.1 \small{\textcolor{teal}{(+2.4)}}} &
  \multicolumn{1}{l|}{31.2 \small{\textcolor{purple}{(-3.7)}}} &
  \multicolumn{1}{l|}{\textbf{32.8} \small{\textcolor{purple}{(-12.9)}}} &
  30.3 \small{\textcolor{purple}{(-14.7)}} \\ 
Rich  &
  \multicolumn{1}{l|}{20.8 \small{\textcolor{purple}{(-2.1)}}} &
  \multicolumn{1}{l|}{0.329 \small{\textcolor{purple}{(-4.4)}}} &
  \multicolumn{1}{l|}{\textbf{41.4} \small{\textcolor{purple}{(-6.6)}}} &
  40.0 \small{\textcolor{purple}{(-5.6)}} \\ 
\end{tabular}%
}
\caption{Average NLLB SigLIP Recall scores for each category and the average difference between default English prompt and country suffix Recall across the four different income groups, grouping country suffixes into income class categories based on their World Bank economic classification. Recall increase is shown in green while Recall drops are highlighted in red. \textit{Best viewed in color.}}
\label{tab:Effect-table}
\end{table}


Interestingly, country suffixes tend to favor image retrieval from income groups that match or are close to their own economic classification. When the four income classes are re-categorized into two classes (lower-income: poor, low-mid and higher-income: up-mid, rich), we find that in 48/63 cases, the image income category with the highest Recall corresponds to the country suffix’s economic class, demonstrating the alignment between income levels of country-suffixes and retrieval performance.



\paragraph{The best-performing country suffixes for lower-income images from a continent are from the same continent.}

We calculate Recall results for lower-income images from 42 countries using the 42 country suffix prompts, yielding a total of 1,764 Recall scores. Using country-suffixes, we group these scores by continent and further categorize them based on the World Bank Income classes of the respective country-suffixes. We present the average Recall and differences compared to default English prompts for each group in \cref{tab:countrytable}.


\begin{table}[h]
\centering
\resizebox{0.9\textwidth}{!}{%
\begin{tabular}{c|c|cccc} 

\multirow{2}{*}{\begin{tabular}[c]{@{}l@{}}\textbf{Image by}\\\textbf{Continent}\end{tabular}} & \multirow{2}{*}{\begin{tabular}[c]{@{}l@{}}\textbf{Country by} \\\textbf{Income}\end{tabular}} & \multicolumn{4}{c}{\textbf{Country Suffix}} \\ 
\cmidrule{3-6}
 &  & \textbf{Africa $\Delta$} & \textbf{America $\Delta$} & \textbf{Asia $\Delta$} & \textbf{Europe $\Delta$} \\ 
\midrule
\multirow{4}{*}{\begin{tabular}[c]{@{}l@{}}Africa\end{tabular}} & Poor & 36.6 \textcolor{teal}{(+15.7)} & 27.2 \textcolor{teal}{(+6.3)} & 23.0 \textcolor{teal}{(+2.1)} & 19.7 \textcolor{purple}{(-1.2)} \\
 & Low-mid & 37.3 \textcolor{teal}{(+10.3)} & 31.2 \textcolor{teal}{(+4.2)} & 25.7 \textcolor{purple}{(-1.4)} & 24.3 \textcolor{purple}{(-2.7)} \\
 & Up-mid & 24.3 \textcolor{teal}{(+2.1)} & 21.6 \textcolor{purple}{(-0.6)} & 17.2 \textcolor{purple}{(-5.0)} & 18.1 \textcolor{purple}{(-4.1)} \\ 

 & \textbf{Average} & {\cellcolor[rgb]{0.933,0.91,0.992}}\textbf{32.7 \textcolor{teal}{(+9.4)}} & \textbf{26.7 \textcolor{teal}{(+3.3)}} & \textbf{22.0} \textcolor{purple}{(-1.4)} & \textbf{20.7} \textcolor{purple}{(-2.7)} \\ 
\midrule
\multirow{4}{*}{\begin{tabular}[c]{@{}l@{}}America\end{tabular}} & Low-mid & 24.5 \textcolor{purple}{(-2.6)} & 26.8 \textcolor{purple}{(-0.4)} & 20.7 \textcolor{purple}{(-6.5)} & 22.4 \textcolor{purple}{(-4.8)} \\
 & Up-mid & 23.4 \textcolor{purple}{(-11.0)} & 35.2 \textcolor{teal}{(+0.9)} & 23.4 \textcolor{purple}{(-10.9)} & 28.9 \textcolor{purple}{(-5.4)} \\
 & Rich & 20.4 \textcolor{purple}{(-15.5)} & 30.4 \textcolor{purple}{(-5.5)} & 22.8 \textcolor{purple}{(-13.1)} & 26.3 \textcolor{purple}{(-9.6)} \\ 

 & \textbf{Average} & \textbf{22.8 }\textcolor{purple}{(-9.7)} & {\cellcolor[rgb]{0.933,0.91,0.992}}\textbf{30.8 \textcolor{purple}{(-1.7)}} & \textbf{22.3} \textcolor{purple}{(-10.2)} & \textbf{25.9} \textcolor{purple}{(-6.6)} \\ 
\midrule
\multirow{4}{*}{\begin{tabular}[c]{@{}l@{}}Asia\end{tabular}} & Low-mid & 29.4 \textcolor{purple}{(-2.4)} & 30.7 \textcolor{purple}{(-1.1)} & 32.7 \textcolor{teal}{(+0.8)} & 27.9 \textcolor{purple}{(-3.9)} \\
 & Up-mid & 28.1 \textcolor{purple}{(-5.1)} & 32.5 \textcolor{purple}{(-0.6)} & 34.6 \textcolor{teal}{(+1.5)} & 30.3 \textcolor{purple}{(-2.8)} \\
 & Rich & 31.0 \textcolor{purple}{(-14.0)} & 33.3 \textcolor{purple}{(-11.7)} & 36.0 \textcolor{purple}{(-9.0)} & 39.6 \textcolor{purple}{(-5.4)} \\ 

 & \textbf{Average} & \textbf{29.5} \textcolor{purple}{(-7.2)} & \textbf{32.2} \textcolor{purple}{(-4.5)} & {\cellcolor[rgb]{0.933,0.91,0.992}}\textbf{34.4 \textcolor{purple}{(-2.2)}} & \textbf{32.6} \textcolor{purple}{(-4.0)} \\ 
\midrule
\multirow{3}{*}{\begin{tabular}[c]{@{}l@{}}Europe\end{tabular}} & Low-mid & 19.6 \textcolor{purple}{(-23.7)} & 29.3 \textcolor{purple}{(-14.0)} & 26.4 \textcolor{purple}{(-16.9)} & 44.3 \textcolor{teal}{(+1.0)} \\
 & Up-mid & 20.4 \textcolor{purple}{(-16.0)} & 26.7 \textcolor{purple}{(-9.7)} & 23.1 \textcolor{purple}{(-13.3)} & 35.9 \textcolor{purple}{(-0.5)} \\ 

 & \textbf{Average} & \textbf{20.0} \textcolor{purple}{(-19.9)} & \textbf{28.0} \textcolor{purple}{(-11.9)} & \textbf{24.8} \textcolor{purple}{(-15.1)} & {\cellcolor[rgb]{0.933,0.91,0.992}}\textbf{40.1 \textcolor{teal}{(+0.3)}} \\
\end{tabular}
}

\caption{Average NLLB SigLIP Recall scores for each continent and the average difference between default English prompt Recall and country suffix Recall from the four continents, grouping lower-income images according to continents and further into groups of countries arranged by income class categories based on their World Bank economic classification. \textit{Best viewed in color.}}
\label{tab:countrytable}
\end{table}

Our findings emphasize the significance of regional specificity in data collection, as the best-performing suffixes align with their respective continents (shown by the diagonal of bold values in \Cref{tab:countrytable}). The results indicate that lower-income images from African nations benefit significantly from including country suffixes. In contrast, data from America and Asia show no Recall improvements, underscoring the necessity for tailored data collection strategies across regions. Notably, lower-income images from African countries exhibit a Recall score of 36.6, reflecting the highest performance increase of 15.7 when using African suffixes. The positive impact of country suffix prompt additions is particularly pronounced for Africa, as the prompts enhance performance on underrepresented data by shifting model inference from its learned standard. This effect is crucial given the current datasets often lack representation from African countries and poor households. Additionally, the similarities among African countries contribute to this improved performance.

Meanwhile, we find no Recall enhancements for higher-income data, regardless of the alignment between images and country suffixes (see \cref{tab:higher_images_country}).





\begin{table}[ht]

\noindent\begin{minipage}{\linewidth}
\centering
\resizebox{0.9\textwidth}{!}{%
\begin{tabular}{c|c|cccc}

\multirow{2}{*}{\begin{tabular}[c]{@{}l@{}}\textbf{Image by}\\\textbf{Continent}\end{tabular}} & \multirow{2}{*}{\begin{tabular}[c]{@{}l@{}}\textbf{Country by} \\\textbf{Income}\end{tabular}} & \multicolumn{4}{c}{\textbf{Country Suffix}} \\ 
\cmidrule{3-6}


&  & \textbf{Africa $\Delta$} & \textbf{America $\Delta$} & \textbf{Asia $\Delta$} & \textbf{Europe $\Delta$} \\ 
\midrule

\multirow{4}{*}{\begin{tabular}[c]{@{}l@{}}Africa\end{tabular}} & Poor & 42.3 \textcolor{purple}{(-1.4)} & 40.3 \textcolor{purple}{(-3.4)} & 36.0 \textcolor{purple}{(-7.7)} & 40.3 \textcolor{purple}{(-3.4)} \\
 & Low-mid & 41.8 \textcolor{purple}{(-8.8)} & 39.7 \textcolor{purple}{(-10.9)} & 37.0 \textcolor{purple}{(-13.6)} & 40.7 \textcolor{purple}{(-9.9)} \\
 & Up-mid & 33.3 \textcolor{purple}{(-5.9)} & 32.4 \textcolor{purple}{(-6.8)} & 25.7 \textcolor{purple}{(-13.5)} & 33.4 \textcolor{purple}{(-5.8)} \\ 

 & \textbf{Average} & {\cellcolor[rgb]{0.933,0.91,0.992}}\textbf{39.1 \textcolor{purple}{(-5.4)}} & \textbf{37.5 }\textcolor{purple}{(-7.0)} & \textbf{32.9} \textcolor{purple}{(-11.6)} & \textbf{38.1} \textcolor{purple}{(-6.4)} \\ 
\hline
\multirow{3}{*}{\begin{tabular}[c]{@{}l@{}} America \end{tabular}} & Up-mid & 24.3 \textcolor{purple}{(-20.8)} & 37.6 \textcolor{purple}{(-7.5)} & 26.9 \textcolor{purple}{(-18.3)} & 37.2 \textcolor{purple}{(-8.0)} \\
 & Rich & 28.7 \textcolor{purple}{(-33.4)} & 45.7 \textcolor{purple}{(-16.4)} & 32.5 \textcolor{purple}{(-29.6)} & 44.7 \textcolor{purple}{(-17.4)} \\ 

 & \textbf{Average} & \textbf{26.5} \textcolor{purple}{(-27.1)} & {\cellcolor[rgb]{0.933,0.91,0.992}}\textbf{41.7 \textcolor{purple}{(-12.0)}} & \textbf{29.7} \textcolor{purple}{(-24.0)} & \textbf{41.0} \textcolor{purple}{(-12.7)} \\ 
\hline
\multirow{4}{*}{\begin{tabular}[c]{@{}l@{}} Asia\end{tabular}} & Low-mid & 33.8 (\textcolor{purple}{-17.2)} & 38.4 \textcolor{purple}{(-12.6)} & 40.4 \textcolor{purple}{(-10.5)} & 43.4 \textcolor{purple}{(-7.6)} \\
 & Up-mid & 30.9 \textcolor{purple}{(-21.0)} & 40.5 \textcolor{purple}{(-11.4)} & 41.2 \textcolor{purple}{(-10.7)} & 46.5 \textcolor{purple}{(-5.4)} \\
 & Rich & 28.2 \textcolor{purple}{(-19.9)} & 36.1 \textcolor{purple}{(-12.0)} & 36.5 \textcolor{purple}{(-11.6)} & 40.1 \textcolor{purple}{(-8.0)} \\ 

 & \textbf{Average} & \textbf{31.0} \textcolor{purple}{(-19.4)} & \textbf{38.3} \textcolor{purple}{(-12.0)} & \textbf{39.4} \textcolor{purple}{(-10.9)} & {\cellcolor[rgb]{0.933,0.91,0.992}}\textbf{43.3 \textcolor{purple}{(-7.0)}} \\ 
\hline
\multirow{4}{*}{\begin{tabular}[c]{@{}l@{}}Europe\end{tabular}} & Low-mid & 22.3 \textcolor{purple}{(-23.9)} & 33.3 \textcolor{purple}{(-13.1)} & 26.6 \textcolor{purple}{(-19.6)} & 42.2 \textcolor{purple}{(-0.4)} \\
 & Up-mid & 18.9 \textcolor{purple}{(-25.0)} & 29.2 \textcolor{purple}{(-14.7)} & 24.5 \textcolor{purple}{(-19.4)} & 40.7 \textcolor{purple}{(-3.2)} \\
 & Rich & 19.0 \textcolor{purple}{(-20.8)} & 27.8 \textcolor{purple}{(-12.0)} & 21.5 \textcolor{purple}{(-18.3)} & 37.6 \textcolor{purple}{(-2.2)} \\ 

 & \textbf{Average} & \textbf{20.1} \textcolor{purple}{(-23.2)} & \textbf{30.0} \textcolor{purple}{(-13.3)} & \textbf{24.2} \textcolor{purple}{(-19.1)} & {\cellcolor[rgb]{0.933,0.91,0.992}}\textbf{40.2 \textcolor{purple}{(-3.1)}} \\

\end{tabular}
}
\end{minipage}
    \caption{Grouping higher income images according to continents and further into groups of countries arranged by income class categories based on their World Bank economic classification, this table shows the average NLLB SigLIP Recall scores for each continent and the average difference between default English prompt Recall and country suffix Recall from the four continents.}
    \label{tab:higher_images_country}
\end{table}

\subsection{RQ3. Does adding income information improve retrieval performance for lower-income images?} \label{rq3}

We create three categories of income suffixes, \textit{poor}, \textit{rich}, and \textit{neutral}, as described in \Cref{income_suffix}. 
We repeat the image retrieval experiments from previous research questions to determine the Recall for images from each topic. 
We group and analyze these results across countries and income groups.

\paragraph{\textit{Poor} income suffixes yield the best performance on most lower-income images.}

Our analysis reveals that the \textit{poor} income suffix prompt achieves the highest performance in 26/42 countries with lower-income images. In 12/42, default English prompts outperform all income suffixes. Nevertheless, most (30/42) countries show Recall improvements when using one of the income suffixes. 

\begin{figure}[h]
    \centering
      \vskip -0.1in
    \includegraphics[width=0.6\textwidth]{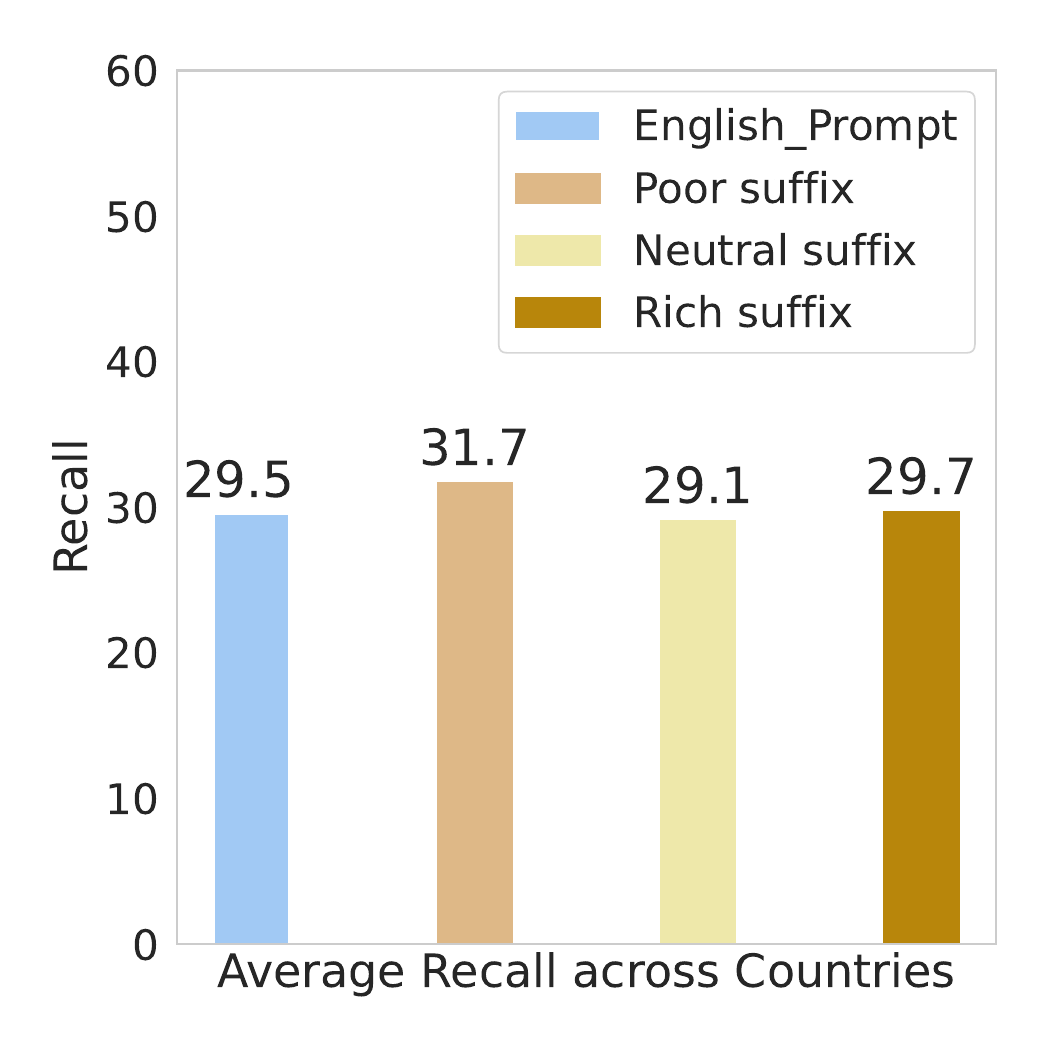}
    \caption{Average Recall with NLLB SigLIP  over \textit{poor} and \textit{lower-middle} income images, for English and Income Suffix prompts. \textit{Best viewed in color.}}
    \vskip -0.15in
    \label{fig:NLLBIncomebest}
\end{figure}

We illustrate in \Cref{fig:NLLBIncomebest} the aggregate the average Recall scores for all 42 countries across default English and the income suffix prompts. 
Notably, the \textit{poor} income suffix demonstrates the best Recall, effectively retrieving a diverse array of images, as shown by the example in \cref{fig:main}.
Recall scores for four sample countries are in Appendix \Cref{fig:NLLBIncomeisbest}.



\paragraph{Images from the \textit{poor} income group benefit the most from income suffixes.}

We group the data into four income groups (by household income) and further categorize them according to the World Bank income classification of their country of origin. In \Cref{tab:incometable}, we show the Recall scores and performance improvements relative to the default English prompts for each data group.

We find that income suffixes predominantly benefit data from poor households and some from low-mid income households, while data from other income groups do not show Recall increases.



\begin{table}[h]
\vskip -0.05in
\resizebox{0.75\textwidth}{!}{%
\begin{tabular}{c|c|ccc}
\multirow{2}{*}{\textbf{\begin{tabular}[c]{@{}l@{}}Images by\\ Income\end{tabular}}} &
  \multirow{2}{*}{\textbf{\begin{tabular}[c]{@{}c@{}}Country\\ by Income\end{tabular}}} &
  \multicolumn{3}{c}{\textbf{Income Suffix}} \\ \cmidrule{3-5} 
 &  & \textbf{Poor $\Delta$} & \textbf{Rich $\Delta$} & \textbf{Neutral $\Delta$} \\ \midrule
\multirow{4}{*}{\begin{tabular}[c]{@{}l@{}}Poor\end{tabular}} & Poor & 26.8 \textcolor{teal}{(+4.4)} & 21.9 \textcolor{purple}{(-0.5)} & 20.7 \textcolor{purple}{(-1.7)} \\
 & Low-mid & 30.0 \textcolor{teal}{(+7.6)} & 26.0 \textcolor{teal}{(+3.6)} & 24.6 \textcolor{teal}{(+2.2)} \\
 & Up-mid & 31.9 \textcolor{teal}{(+9.5)} & 28.0 \textcolor{teal}{(+5.6)} & 26.3 \textcolor{teal}{(+3.9)} \\  
 & \textbf{Average} & \textbf{29.6} \textcolor{teal}{(+7.2)} & \textbf{25.3} \textcolor{teal}{(+2.9)} & \textbf{23.9} \textcolor{teal}{(+1.5)}\\ \midrule
\multirow{5}{*}{\begin{tabular}[c]{@{}l@{}}Low-mid\end{tabular}} & Poor & 33.3 \textcolor{purple}{(-2.6)} & 33.3 \textcolor{purple}{(-2.6)} & 33.5 \textcolor{purple}{(-2.4)} \\
 & Low-mid & 36.5 \textcolor{teal}{(+0.6)} & 35.6 \textcolor{purple}{(-0.3)} & 35.7 \textcolor{purple}{(-0.2)} \\
 & Up-mid & 30.6 \textcolor{purple}{(-5.3)} & 30.5 \textcolor{purple}{(-5.4)} & 30.3 \textcolor{purple}{(-5.6)} \\
 & Rich & 35.5 \textcolor{purple}{(-0.5)} & 38.1 \textcolor{teal}{(+2.2)} & 37.2 \textcolor{teal}{(+1.3)} \\  
 & \textbf{Average} & \textbf{34.0} \textcolor{purple}{(-2.0)} & \textbf{34.4} \textcolor{purple}{(-1.5)} & \textbf{37.2} \textcolor{purple}{(-1.7)} \\ \midrule
\multirow{5}{*}{\begin{tabular}[c]{@{}l@{}}Up-mid\end{tabular}} & Poor & 31.4 \textcolor{purple}{(-14.2)} & 36.4 \textcolor{purple}{(-9.2)} & 32.8 \textcolor{purple}{(-12.8)} \\
 & Low-mid & 37.6 \textcolor{purple}{(-8.1)} & 42.4 \textcolor{purple}{(-3.2)} & 44.5 \textcolor{purple}{(-1.1)} \\
 & Up-mid & 33.0 \textcolor{purple}{(-12.6)} & 38.0 \textcolor{purple}{(-7.6)} & 41.3 \textcolor{purple}{(-4.3)} \\
 & Rich & 28.4 \textcolor{purple}{(-17.4)} & 33.4 \textcolor{purple}{(-12.2)} & 36.3 \textcolor{purple}{(-9.3)} \\ 
 &  \textbf{Average} & \textbf{32.6} \textcolor{purple}{(-13.1)} & \textbf{37.6} \textcolor{purple}{(-8.1)} & \textbf{38.7} \textcolor{purple}{(-6.9)} \\ \midrule
\multirow{5}{*}{\begin{tabular}[c]{@{}l@{}}Rich \end{tabular}} & Poor & 29.6 \textcolor{purple}{(-14.0)} & 42.6 \textcolor{purple}{(-1.0)} & 42.4 \textcolor{purple}{(-1.2)} \\
 & Low-mid & 30.6 \textcolor{purple}{(-13.0)} & 38.0 \textcolor{purple}{(-5.6)} & 38.5 \textcolor{purple}{(-5.1)} \\
 & Up-mid & 25.8 \textcolor{purple}{(-17.8)} & 33.8 \textcolor{purple}{(-9.8)} & 36.1 \textcolor{purple}{(-7.5)} \\
 & Rich & 25.5 \textcolor{purple}{(-18.1)} & 33.8 \textcolor{purple}{(-9.8)} & 36.5 \textcolor{purple}{(-7.1)} \\  
 & \textbf{Average} & \textbf{27.9} \textcolor{purple}{(-15.7)} & \textbf{37.1} \textcolor{purple}{(-6.6)} & \textbf{38.4} \textcolor{purple}{(-5.2)} \\ 
\end{tabular}%
}
\caption{Average NLLB SigLIP Recall scores for each category and the average difference between default English prompt Recall and income suffix recall, grouping images according to household income level and separating countries into income class categories based on their World Bank economic classification.}
\label{tab:incometable}
\end{table}

An interesting finding is that all income suffixes, including \textit{rich} and \textit{neutral}, result in decreased Recall for higher-income images (i.e., \textit{up-mid} and \textit{rich}). This suggests that default English prompts yield the best results for higher-income images, likely due to their high representation in AI models and datasets as the "standard representation." Consequently, the inclusion of socioeconomic status information may lead the model to prioritize lower-income images over higher-income ones. This phenomenon is evident in the results, which show Recall improvements for lower-income images while diminishing Recall for higher-income images, potentially indicating a shift in the model's perspective away from its default understanding of the topic.

\begin{table}[ht]
\centering
\resizebox{\columnwidth}{!}{%
\begin{tabular}{l|lllll}

Model              &  & \multicolumn{4}{c}{Suffix Prompts} \\ \cline{2-6} 
 & English & Country & Poor & Rich & Neutral \\ \hline
NLLB SigLIP                 & 30.3      & 41.6      & 32.1      & 30.2     & 29.5     \\
Sentence Transformers MCLIP & 22.2      & 44.9      & 28.6      & 26.7     & 25.2     \\
Open AI CLIP ViT 32/B       & 25.4      & 45.0      & 30.0      & 28.1     & 28.3    
\end{tabular}%
}
\caption{Average Recall over lower-income images across 39 countries for English, Country Suffix, and three Income Suffix prompts for three LMM models}
\label{tab:threeLMMmodels}
\end{table}

\begin{table}[ht]
\centering
\resizebox{\columnwidth}{!}{%
\begin{tabular}{l|ll}
Model                       & English & \multicolumn{1}{c}{Native Translated} \\ \hline
NLLB SigLIP                 & 31.3    & 28.5                                  \\
Sentence Transformers MCLIP & 24.9    & 18.3                                 
\end{tabular}%
}
\caption{Average Recall over lower-income images across 25 countries for English and native translated language prompts for two multilingual LMM models.}
\label{tab:country_sta}
\end{table}


\subsection{Results Significance and Generalizability}
We conducted the Wilcoxon Signed Rank \cite{woolson2005wilcoxon} test (p-value < 0.05) to assess the statistical significance of our findings. The results indicated that the differences between the default English prompt results and each prompt intervention were statistically significant, except for the 'rich' and 'neutral' income suffix prompts (more details in Appendix \cref{tab:wilcoxon}). 

Although our primary focus is on the NLLB-CLIP-SigLIP results, we confirmed that these findings are consistent across the two other LMM models we tested (Open AI's CLIP ViT B/32 \cite{Radford2021LearningTV} and Sentence Transformers clip-ViT-B-32-multilingual-v1 \cite{reimers-2019-sentence-bert}). A summary of results from these additional models is included in \cref{tab:threeLMMmodels} and \cref{tab:country_sta}.

\section{Lessons Learned}

We highlight key insights learned from our findings and present them below.

\paragraph{Current multilingual LMM models do not significantly improve diversity and representation.}

Our results from Section \labelcref{rq1} demonstrate that English prompts perform better on lower (and higher) income images than prompts translated to a non-English language widely spoken in the region where the data was collected. Since the quality of translations, quantity of training data available for these languages, and consequently, the performance of AI models in these languages is lower than that of English, these findings are not very surprising. We can look forward to better non-English language performance as multilingual LMM models improve.

\paragraph{Location and socio-economic attributes improve retrieval performance for lower-income images.} 

We find that adding geographical and socioeconomic attributes (including rich and neutral attributes) to prompts leads to an increased model preference for lower-income images over higher-income images, as demonstrated in Section \labelcref{rq2}. Images from poor households typically suffer the most from underrepresentation as they differ the most from the type of images available on the internet \cite{roslingfactfulness}.
Since LMM models have learned representations from high-income images as the standard, then adding more information to the prompt (such as country suffixes like 'Malawi', income suffixes describing poverty or wealth, or neutral suffixes like 'a place') shifts the perspective of the model to retrieve images that are more diverse and less contained to the learned 'standard'.

\paragraph{Images with less standard topic appearances are retrieved using income suffix and country suffix prompts.}

Inspection of the retrieved images reveals that images with topic appearances commonly found in lower-income households previously not retrieved by the default English prompts are being retrieved with these prompts as shown in \Cref{fig:main}.
For example, \textit{pit latrines} and \textit{forest-style toilets} previously left out by the default English prompts are retrieved using country suffixes (\textit{Burundi} and \textit{Cameroon}) and the \textit{poor} income suffixes. Another example is ``leaves'' as ``toilet paper'' retrieved by \textit{Liberia} and \textit{Cameroon} country suffixes but excluded by the default English prompt.

\section{Conclusion}

In this paper, we addressed the uneven performance of LMM models across different countries and income levels. We explored three attribute-integrated prompting strategies: (1) translation of text prompts to native non-English languages, (2) addition of geographic information, and (3) addition of socioeconomic attributes. 
We found that integrating geographical and socioeconomic information into prompts enhances LMM model performance on images from lower-income households and retrieves more diverse label representations.
Furthermore, we identified and highlighted the contexts where the proposed prompting techniques work best and shared our insights to improve representation in LMM models and datasets.
Our code can be used to evaluate the performance of other LMM  models and datasets and is publicly available at \href{https://github.com/Anniejoan/Uplifting-Lower-income-data}{Analysis for Uplifting lower-income data}.

\section*{Limitations}

\paragraph{Translation Quality}
 We note that, while NLLB-200-distilled-600M is reputed as a SOTA machine translation model, it does not have perfect accuracy on machine translation across all the languages it supports. We acknowledge that the quality of translations obtained from NLLB-200-distilled-600M greatly impacts our results.

\paragraph{Data Coverage}
Our study is constrained by the reach of the Dollar Street dataset and the number of contributions obtained from each region. Therefore, we do not account for data from other regions that are not included in the dataset.

\paragraph{Choice of Attributes}
We acknowledge that other attributes (e.g., physical attributes like color and material) of the objects in the images could be integrated into prompts to improve performance. However, we choose to focus on geographic and socioeconomic attributes since they are broad enough to include all possible topic appearances related to that attribute and their impact on data belonging to different countries and income groups can measured directly.

\paragraph{Diverse Data Availability}
While our methods facilitate the improvement of diversity during dataset annotation, these strategies cannot overcome the representation issues within the actual pool of images available for annotation.



\section*{Ethics Statement}

Through this work, we aim to contribute toward improving diversity in AI models and even out the disparate impact of these models on the public, especially on underrepresented groups. The strategies discussed in our work can be used to prioritize the retrieval of lower-income images for balancing skewed data representation or domain-specific applications in AI.  However, we do not encourage the use of these strategies to promote over-representation or the inclusion of one group over another in contexts that affect all members of the general public. 

Our decision to use the NLLB-SigLIP model exemplifies our commitment to inclusive models that benefit as many people as possible, especially underrepresented groups.
While researching technologically advanced communities is easier and less resource-intensive, we stress the importance of making AI design decisions that do not exclude communities with limited access to technology.

\section*{Acknowledgements}
We are grateful to the Language and Information Technologies (LIT) lab members at the University of Michigan for their insightful discussions and feedback during the project's early stages. 
This project was partially funded by a grant from the Department of State (\#STC10023GR0014). Any opinions, findings, conclusions, or recommendations expressed in this material are those of the authors and do not necessarily reflect the views of the Department of State.

\bibliography{anthology,custom}

\appendix

\section{Appendix}
\label{sec:appendix}
 \subsection{Non-English Languages}
 We use the following non-English languages in our experiments. \textit{'German', 'Spanish', 'Portuguese', 'French', 'Chinese', 'Czech', 'Danish', 'Arabic', 'Hindi', 'Indonesian', 'Farsi-Persian', 'Italian', 'Russian', 'Mongolian', 'Burmese', 'Dutch', 'Urdu', 'Romanian', 'Serbian', 'Korean', 'Swedish', 'Thai', 'Turkish', 'Ukrainian', 'Vietnamese', 'Bengali', 'Khmer', 'Oromo', 'Ewe', 'Creole', 'Swahili', 'Nepali', 'Hausa', 'Kyrgyz', 'Tagalog', 'Kinyarwanda', 'Somali', 'Zulu', 'Sinhala', 'Shona'}


\begin{table*}[]
\centering
\resizebox{\columnwidth}{!}{%
\begin{tabular}{|lllll|}
\hline
\multicolumn{1}{|l|}{Countries} &
  \multicolumn{1}{c|}{\begin{tabular}[c]{@{}c@{}}Non-English \\ Language\end{tabular}} &
  \multicolumn{1}{c|}{\begin{tabular}[c]{@{}l@{}}Image Income \\Classes\end{tabular}} &
  \multicolumn{1}{l|}{\begin{tabular}[c]{@{}l@{}}World Bank\\ Country Economic Classes\end{tabular}} &
  Continent \\ \hline
Austria          & German            & Rich, Up-mid                & High     & Europe  \\
Bangladesh       & Bengali           & Poor                        & Low-mid  & Asia    \\
Bolivia          & Spanish           & Low-mid, poor               & Low-mid  & America \\
Brazil           & Portuguese        & Rich, Up-mid, Low-mid       & Up-mid   & America \\
Burkina Faso     & French            & Poor                        & Poor/low & Africa  \\
Burundi          & French            & Poor                        & Poor/low & Africa  \\
Cambodia         & Khmer             & Up-mid, Low-mid, Poor       & Low-mid  & Asia    \\
Cameroon         & French            & Up-mid, Low-mid, Poor       & Low-mid  & Africa  \\
Canada           & French            & Rich                        & High     & America \\
China            & Chinese           & Rich, Up-mid, Low-mid, Poor & Up-mid   & Asia    \\
Colombia         & Spanish           & Rich, Up-mid, Low-mid, Poor & Up-mid   & America \\
Cote d'Ivoire    & French            & Poor                        & Low-mid  & Africa  \\
Czech Republic   & Czech             & Rich                        & High     & Europe  \\
Denmark          & Danish            & Rich                        & High     & Europe  \\
Egypt            & Arabic            & Up-mid                      & Low-mid  & Africa  \\
Ethiopia         & Oromo             & Rich, Up-mid, Low-mid       & Poor/low & Africa  \\
France           & French            & Rich, Up-mid                & High     & Europe  \\
Ghana            & Ewe               & Low-mid                     & Low-mid  & Africa  \\
Guatemala        & Spanish           & Low-mid                     & Up-mid   & America \\
Haiti            & Creole            & Poor                        & Low-mid  & America \\
India            & Hindi             & Rich, Up-mid, Low-mid, Poor & Low-mid  & Asia    \\
Indonesia        & Bahasa Indonesian & Rich, Up-mid, Low-mid, Poor & Up-mid   & Asia    \\
Iran             & Farsi (Persian)   & Rich, Up-mid                & Low-mid  & Asia    \\
Italy            & Italian           & Rich                        & High     & Europe  \\
Jordan           & Arabic            & Rich, Low-mid               & Low-mid  & Asia    \\
Kazakhstan       & Russian           & Up-mid                      & Up-mid   & Asia    \\
Kenya            & Swahili           & Rich, Low-mid, Poor         & Low-mid  & Africa  \\
Kyrgyzstan       & Kyrgyz            & Up-mid                      & Low-mid  & Asia    \\
Lebanon          & Arabic            & Up-mid                      & Low-mid  & Asia    \\
Liberia          & -                 & Poor                        & Poor/low & Africa  \\
Malawi           & -                 & Poor                        & Poor/low & Africa  \\
Mexico           & Spanish           & Rich, Up-mid                & Up-mid   & America \\
Mongolia         & Mongolian         & Low-mid                     & Low-mid  & Asia    \\
Myanmar          & Burmese           & Low-mid, poor               & Low-mid  & Asia    \\
Nepal            & Nepali            & Rich, Up-mid, Low-mid, Poor & Low-mid  & Asia    \\
Netherlands      & Dutch             & Rich, Up-mid                & High     & Europe  \\
Nigeria          & Hausa             & Rich, Up-mid, Low-mid, Poor & Low-mid  & Africa  \\
Pakistan         & Urdu              & Rich, Up-mid, Low-mid, Poor & Low-mid  & Asia    \\
Palenstine       & Arabic            & Low-mid, poor               & Low-mid  & Asia    \\
Papua New Guinea & -                 & Poor                        & Low-mid  & Asia    \\
Peru             & Spanish           & Low-mid, poor               & Up-mid   & America \\
Philippines      & Tagalog           & Up-mid, Low-mid, Poor       & Low-mid  & Asia    \\
Romania          & Romanian          & Rich                        & High     & Europe  \\
Russia           & Russian           & Rich, Up-mid                & Up-mid   & Europe  \\
Rwanda           & Kinyarwanda       & Low-mid, poor               & Poor/low & Africa  \\
Serbia           & Serbian           & Rich, Up-mid, Low-mid       & Up-mid   & Europe  \\
Somalia          & Somali            & Poor                        & Poor/low & Africa  \\
South Africa     & Zulu              & Rich, Up-mid, Low-mid, Poor & Up-mid   & Africa  \\

 \\ \hline
\end{tabular}%
}
\label{tab:country_stats}
\end{table*}

\begin{table*}[]
\centering
\resizebox{\columnwidth}{!}{%
\begin{tabular}{|lllll|}
\hline

\multicolumn{1}{|l|}{Countries} &
  \multicolumn{1}{c|}{\begin{tabular}[c]{@{}c@{}}Non-English \\ Language\end{tabular}} &
  \multicolumn{1}{c|}{\begin{tabular}[c]{@{}l@{}}Image Income\\ Classes\end{tabular}} &
  \multicolumn{1}{l|}{\begin{tabular}[c]{@{}l@{}}World Bank\\ Country Economic Classes\end{tabular}} &
  Continent \\ \hline

  South Korea      & Korean            & Rich, Up-mid, Low-mid       & High     & Asia    \\
Spain            & Spanish           & Rich                        & High     & Europe  \\
Sri Lanka        & Sinhala           & Up-mid                      & Low-mid  & Asia    \\
Sweden           & Swedish           & Rich, Up-mid                & High     & Europe  \\
Switzerland      & German            & Rich                        & High     & Europe  \\
Tanzania         & Swahili           & Up-mid, Low-mid, Poor       & Low-mid  & Africa  \\
Thailand         & Thai              & Up-mid, Low-mid, Poor       & Up-mid   & Asia    \\
Togo             & French            & Low-mid, poor               & Poor/low & Africa  \\
Tunisia          & Arabic            & Low-mid, poor               & Low-mid  & Africa  \\
Turkey           & Turkish           & Rich                        & Up-mid   & Europe  \\
Ukraine          & Ukrainian         & Rich, Up-mid, Low-mid       & Low-mid  & Europe  \\
United Kingdom   & -                 & Rich, Up-mid                & High     & Europe  \\
United States    & Spanish           & Rich, Up-mid, Low-mid       & High     & America \\
Vietnam          & Vietnamese        & Low-mid, Rich               & Low-mid  & Asia    \\
Zimbabwe         & Shona             & Poor                        & Low-mid  & Africa  \\ \hline
\end{tabular}%
}
\caption{Table displaying the 63 Dollar Street countries, their major non-English language, income levels of contributions for that country, World Bank income class, and their continent.}
\label{tab:country_stats2}
\end{table*}

\begin{figure}[h]
    \centering
    \includegraphics[width=\textwidth]{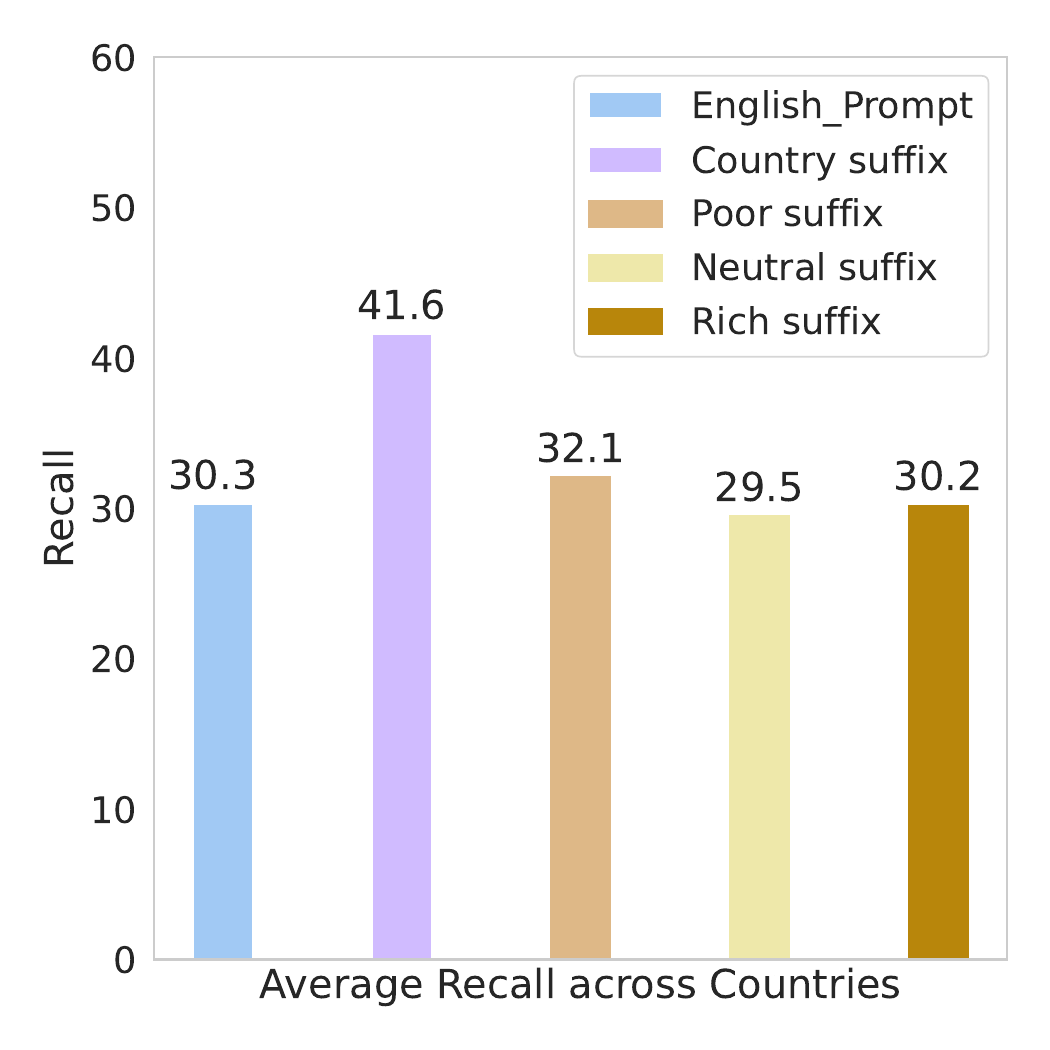}
    \caption{Average Recall over lower-income images across 39 countries for English, Country Suffix, and Income Suffix prompts}
    \label{fig:all_prompts}
\end{figure}


\begin{table}[ht]
\centering
\resizebox{\columnwidth}{!}{%
\begin{tabular}{|l|rrrr|}
\hline
\multirow{2}{*}{\textbf{Languages}} &
  \multicolumn{4}{c|}{\textbf{Income level of images}} \\ \cline{2-5} 
 & \multicolumn{1}{l}{\textbf{Poor}} & \multicolumn{1}{l}{\textbf{Low-mid}} & \multicolumn{1}{l}{\textbf{Up-mid}} & \multicolumn{1}{l|}{\textbf{Rich}} \\ \hline
Arabic &
  \multicolumn{1}{l}{21.0 \small{\textcolor{purple}{(-2.4)}}} &
  \multicolumn{1}{l}{32.0 \small{\textcolor{purple}{(-3.9)}}} &
  \multicolumn{1}{l}{38.5 \small{\textcolor{purple}{(-7.1)}}} &
  \multicolumn{1}{l|}{37.3 \small{\textcolor{purple}{(-6.3)}}} \\
Bengali &
  \multicolumn{1}{l}{20.9 \small{\textcolor{purple}{(-1.5)}}} &
  \multicolumn{1}{l}{33.0 \small{\textcolor{purple}{(-2.9)}}} &
  \multicolumn{1}{l}{40.7 \small{\textcolor{purple}{(-4.9)}}} &
  \multicolumn{1}{l|}{38.9 \small{\textcolor{purple}{(-4.7)}}} \\
Burmese &
  \multicolumn{1}{l}{21.5 \small{\textcolor{purple}{(-0.9)}}} &
  \multicolumn{1}{l}{30.4 \small{\textcolor{purple}{(-5.5)}}} &
  \multicolumn{1}{l}{36.0 \small{\textcolor{purple}{(-9.6)}}} &
  \multicolumn{1}{l|}{34.2 \small{\textcolor{purple}{(-9.4)}}} \\
Chinese &
  \multicolumn{1}{l}{21.5 \small{\textcolor{purple}{(-0.9)}}} &
  \multicolumn{1}{l}{32.5 \small{\textcolor{purple}{(-3.4)}}} &
  \multicolumn{1}{l}{39.1 \small{\textcolor{purple}{(-6.5)}}} &
  \multicolumn{1}{l|}{37.3 \small{\textcolor{purple}{(-6.3)}}} \\
Creole &
  \multicolumn{1}{l}{21.0 \small{\textcolor{purple}{(-1.4)}}} &
  \multicolumn{1}{l}{32.6 \small{\textcolor{purple}{(-3.3)}}} &
  \multicolumn{1}{l}{40.1 \small{\textcolor{purple}{(-5.5)}}} &
  \multicolumn{1}{l|}{38.0 \small{\textcolor{purple}{(-5.6)}}} \\
Czech &
  \multicolumn{1}{l}{19.9 \small{\textcolor{purple}{(-2.5)}}} &
  \multicolumn{1}{l}{32.3 \small{\textcolor{purple}{(-3.6)}}} &
  \multicolumn{1}{l}{40.3 \small{\textcolor{purple}{(-5.3)}}} &
  \multicolumn{1}{l|}{38.8 \small{\textcolor{purple}{(-4.8)}}} \\
Danish &
  \multicolumn{1}{l}{20.8 \small{\textcolor{purple}{(-1.6)}}} &
  \multicolumn{1}{l}{33.7 \small{\textcolor{purple}{(-2.2)}}} &
  \multicolumn{1}{l}{41.7 \small{\textcolor{purple}{(-3.9)}}} &
  \multicolumn{1}{l|}{40.0 \small{\textcolor{purple}{(-3.6)}}} \\
Dutch &
  \multicolumn{1}{l}{21.1 \small{\textcolor{purple}{(-1.3)}}} &
  \multicolumn{1}{l}{33.6 \small{\textcolor{purple}{(-2.3)}}} &
  \multicolumn{1}{l}{42.5 \small{\textcolor{purple}{(-3.1)}}} &
  \multicolumn{1}{l|}{40.8 \small{\textcolor{purple}{(-2.8)}}} \\
Ewe &
  14.7 \small{\textcolor{purple}{(-7.7)}} &
  19.3 \small{\textcolor{purple}{(-16.6)}} &
  22.1 \small{\textcolor{purple}{(-23.5)}} &
  20.6 \small{\textcolor{purple}{(-23.0)}} \\
Farsi-Persian &
  21.9 \small{\textcolor{purple}{(-0.5)}} &
  31.8 \small{\textcolor{purple}{(4.1)}} &
  39.1 \small{\textcolor{purple}{(-6.5)}} &
  38.1 \small{\textcolor{purple}{(-5.5)}} \\
French &
  21.7 \small{\textcolor{purple}{(-0.7)}} &
  33.7 \small{\textcolor{purple}{(-2.2)}} &
  42.6 \small{\textcolor{purple}{(-3.0)}} &
  41.4 \small{\textcolor{purple}{(-2.2)}} \\
German &
  21.5 \small{\textcolor{purple}{(-0.9)}} &
  33.1 \small{\textcolor{purple}{(-2.8)}} &
  41 \small{\textcolor{purple}{(-4.6)}} &
  39 \small{\textcolor{purple}{(-4.6)}} \\
Hausa &
  20.6 \small{\textcolor{purple}{(-1.8)}} &
  31.6 \small{\textcolor{purple}{(-4.3)}} &
  38.4 \small{\textcolor{purple}{(-7.2)}} &
  36.4 \small{\textcolor{purple}{(-7.2)}} \\
Hindi &
  22.2 \small{\textcolor{purple}{(-0.2)}} &
  34.5 \small{\textcolor{purple}{(-1.4)}} &
  41.8 \small{\textcolor{purple}{(-3.8)}} &
  40 \small{\textcolor{purple}{(-3.6)}} \\
Indonesian &
      22.1 \small{\textcolor{purple}{(-0.3)}} &
      34.8 \small{\textcolor{purple}{(-1.1)}} &
  42.4 \small{\textcolor{purple}{(-3.2)}} &
  40.5 \small{\textcolor{purple}{(-3.1)}} \\
Italian &
  21.1 \small{\textcolor{purple}{(-1.3)}} &
  34.3 \small{\textcolor{purple}{(-1.6)}} &
  42.7 \small{\textcolor{purple}{(-2.9)}} &
  41.4 \small{\textcolor{purple}{(-2.2)}} \\
Khmer &
  16.4 \small{\textcolor{purple}{(-6.0)}} &
  22.0 \small{\textcolor{purple}{(-13.9)}} &
  24.8 \small{\textcolor{purple}{(-20.8)}} &
  23.2 \small{\textcolor{purple}{(-20.4)}} \\
Kinyarwanda &
  16.7 \small{\textcolor{purple}{(-5.7)}} &
  23.7 \small{\textcolor{purple}{(-12.2)}} &
  28.8 \small{\textcolor{purple}{(-16.8)}} &
  27.2 \small{\textcolor{purple}{(-16.4)}} \\
Korean &
  20.4 \small{\textcolor{purple}{(-2.0)}} &
  32.2 \small{\textcolor{purple}{(-3.7)}} &
  40.2 \small{\textcolor{purple}{(-5.4)}} &
  37.9 \small{\textcolor{purple}{(-5.7)}} \\
Kyrgyz &
  21.6 \small{\textcolor{purple}{(-0.8)}} &
  30.9 \small{\textcolor{purple}{(-5.0)}} &
  36.7 \small{\textcolor{purple}{(-8.9)}} &
  35.7 \small{\textcolor{purple}{(-7.9)}} \\
Mongolian &
  13.7 \small{\textcolor{purple}{(-8.7)}} &
  20.9 \small{\textcolor{purple}{(-1.5)}} &
  25 \small{\textcolor{purple}{(-20.6)}} &
  23.4 \small{\textcolor{purple}{(20.2)}} \\
Nepali &
  20.7 \small{\textcolor{purple}{(-1.7)}} &
  32.5 \small{\textcolor{purple}{(-3.4)}} &
  40.9 \small{\textcolor{purple}{(-4.7)}} &
  39.7 \small{\textcolor{purple}{(-3.9)}} \\
Oromo &
  15.8 \small{\textcolor{purple}{(-6.6)}} &
  20.9 \small{\textcolor{purple}{(-15.0)}} &
  24.7 \small{\textcolor{purple}{(-20.9)}} &
  23.4 \small{\textcolor{purple}{(20.2)}} \\
Portuguese &
  21.3 \small{\textcolor{purple}{(-1.1)}} &
  34 \small{\textcolor{purple}{(-1.9)}} &
  42.6 \small{\textcolor{purple}{(-3.0)}} &
  41.2 \small{\textcolor{purple}{(-2.4)}} \\
Romanian &
  20.3 \small{\textcolor{purple}{(-2.1)}} &
  32.9 \small{\textcolor{purple}{(-3.0)}} &
  41.0 \small{\textcolor{purple}{(-4.6)}} &
  38.9 \small{\textcolor{purple}{(-4.7)}} \\
Russian &
  21.1 \small{\textcolor{purple}{(-1.3)}} &
  33.4 \small{\textcolor{purple}{(-2.5)}} &
  41.5 \small{\textcolor{purple}{(-4.1)}} &
  39.9 \small{\textcolor{purple}{(-3.7)}} \\
Serbian &
  19.2 \small{\textcolor{purple}{(-3.2)}} &
  30.8 \small{\textcolor{purple}{(-5.1)}} &
  37.2 \small{\textcolor{purple}{(-8.4)}} &
  35.6 \small{\textcolor{purple}{(-8.0)}} \\
Shona &
  19.1 \small{\textcolor{purple}{(-3.3)}} &
  27.2 \small{\textcolor{purple}{(-8.7)}} &
  32.2 \small{\textcolor{purple}{(-13.4)}} &
  30.5 \small{\textcolor{purple}{(-13.1)}} \\
Sinhala &
  20.4 \small{\textcolor{purple}{(-2.0)}} &
  32.0 \small{\textcolor{purple}{(-3.9)}} &
  37.9 \small{\textcolor{purple}{(-7.7)}} &
  35.7 \small{\textcolor{purple}{(-7.9)}} \\
Somali &
  19.0 \small{\textcolor{purple}{(-3.4)}} &
  28.5 \small{\textcolor{purple}{(-7.4)}} &
  33.8 \small{\textcolor{purple}{(-11.8)}} &
  31.4 \small{\textcolor{purple}{(-12.2)}} \\
Spanish &
  20.7 \small{\textcolor{purple}{(-1.7)}} &
  33.8 \small{\textcolor{purple}{(-2.1)}} &
  42.5 \small{\textcolor{purple}{(-3.1)}} &
  40.9 \small{\textcolor{purple}{(-2.7)}} \\
Swahili &
  22.1 \small{\textcolor{purple}{(-0.3)}} &
  33.6 \small{\textcolor{purple}{(-2.3)}} &
  41.3 \small{\textcolor{purple}{(-4.3)}} &
  38.9 \small{\textcolor{purple}{(-4.7)}} \\
Swedish &
  20.5 \small{\textcolor{purple}{(-1.9)}} &
  33.0 \small{\textcolor{purple}{(-2.9)}} &
  40.5 \small{\textcolor{purple}{(-5.1)}} &
  38.6 \small{\textcolor{purple}{(-5.0)}} \\
Tagalog &
  21.4 \small{\textcolor{purple}{(-1.0)}} &
  33.2 \small{\textcolor{purple}{(-2.7)}} &
  39.4 \small{\textcolor{purple}{(-6.2)}} &
  37.5 \small{\textcolor{purple}{(-6.1)}} \\
Thai &
  19.7 \small{\textcolor{purple}{(-2.7)}} &
  29.7 \small{\textcolor{purple}{(-6.2)}} &
  34.9 \small{\textcolor{purple}{(-10.7)}} &
  33.6 \small{\textcolor{purple}{(-10.0)}} \\
Turkish &
  20.5 \small{\textcolor{purple}{(-1.9)}} &
  31.6 \small{\textcolor{purple}{(-4.3)}} &
  39.5 \small{\textcolor{purple}{(-6.1)}} &
  38.5 \small{\textcolor{purple}{(-5.1)}} \\
Ukrainian &
  20.7 \small{\textcolor{purple}{(-1.7)}} &
  33.0 \small{\textcolor{purple}{(-2.9)}} &
  40.7 \small{\textcolor{purple}{(-4.9)}} &
  38.7 \small{\textcolor{purple}{(-4.9)}} \\
Urdu &
  21.5 \small{\textcolor{purple}{(-0.9)}} &
  33.0 \small{\textcolor{purple}{(-2.9)}} &
  40.6 \small{\textcolor{purple}{(-5.0)}} &
  39.1 \small{\textcolor{purple}{(-4.5)}} \\
Vietnamese &
  20.6 \small{\textcolor{purple}{(-1.8)}} &
  32.8 \small{\textcolor{purple}{(-3.1)}} &
  41.1 \small{\textcolor{purple}{(-4.5)}} &
  39.5 \small{\textcolor{purple}{(-4.1)}} \\
Zulu &
  19.9 \small{\textcolor{purple}{(-2.5)}} &
  29.9 \small{\textcolor{purple}{(-6.0)}} &
  35.2 \small{\textcolor{purple}{(-10.4)}} &
  33.4 \small{\textcolor{purple}{(-10.2)}} \\ \hline
\end{tabular}%
}
\caption{Non-English prompts lead to a decrease in Recall scores across all income levels. Table of the differences (rounded to 1 d.p.) between Recall scores for non-English language prompts and Recall scores for default English prompts for all data grouped into income levels. }
\label{tab:drop_dom_lang_all}
\end{table}


\begin{table*}[ht]
\centering
\resizebox{\columnwidth}{!}{%
\begin{tabular}{lllll}
\hline
\multicolumn{1}{|l|}{} &
  \multicolumn{4}{c|}{\textbf{Income levels}} \\ \cline{2-5} 
\multicolumn{1}{|l|}{\multirow{-2}{*}{\textbf{Country Suffix}}} &
  \textbf{Poor \hspace{2pt}  $\Delta$} &
  \textbf{Low-mid \hspace{2pt}  $\Delta$} &
  \textbf{Up-mid \hspace{2pt}  $\Delta$} &
  \multicolumn{1}{l|}{\textbf{Rich \hspace{2pt}  $\Delta$}} \\
\rowcolor[HTML]{FBDDF5} 
Burkina Faso &
  \textbf{0.327} \small{\textcolor{teal}{ (+0.103)}} &
  0.303 \small{\textcolor{purple}{(-0.056)}} &
  0.227 \small{\textcolor{purple}{(-0.229)}} &
  0.185 \small{\textcolor{purple}{(-0.251)}} \\
\rowcolor[HTML]{FBDDF5} 
Burundi &
  \textbf{0.331} \small{\textcolor{teal}{ (+0.107)}} &
  0.279 \small{\textcolor{purple}{(-0.08)}} &
  0.197 \small{\textcolor{purple}{(-0.259)}} &
  0.154 \small{\textcolor{purple}{(-0.282)}} \\
\rowcolor[HTML]{FBDDF5} 
Ethiopia &
  \textbf{0.334} \small{\textcolor{teal}{ (+0.11)}} &
  0.313 \small{\textcolor{purple}{(-0.046)}} &
  0.269 \small{\textcolor{purple}{(-0.187)}} &
  0.227 \small{\textcolor{purple}{(-0.209)}} \\
\rowcolor[HTML]{FBDDF5} 
Liberia &
  \textbf{0.327} \small{\textcolor{teal}{ (+0.103)}} &
  0.303 \small{\textcolor{purple}{(-0.056)}} &
  0.249 \small{\textcolor{purple}{(-0.207)}} &
  0.205 \small{\textcolor{purple}{(-0.231)}} \\
\rowcolor[HTML]{FBDDF5} 
Malawi &
  0.301 \small{\textcolor{teal}{(+0.077)}} &
  \textbf{0.32} \small{\textcolor{purple}{(-0.39)}} &
  0.286 \small{\textcolor{purple}{(-0.17)}} &
  0.254 \small{\textcolor{purple}{(-0.182)}} \\
\rowcolor[HTML]{FBDDF5} 
Rwanda &
  \textbf{0.334} \small{\textcolor{teal}{ (+0.11)}} &
  0.322 \small{\textcolor{purple}{(-0.037)}} &
  0.249 \small{\textcolor{purple}{(-0.207)}} &
  0.204 \small{\textcolor{purple}{(-0.232)}} \\
\rowcolor[HTML]{FBDDF5} 
Somalia &
  \textbf{0.318} \small{\textcolor{teal}{ (+0.094)}} &
  0.296 \small{\textcolor{purple}{(-0.063)}} &
  0.243 \small{\textcolor{purple}{(-0.213)}} &
  0.209 \small{\textcolor{purple}{(-0.227)}} \\
\rowcolor[HTML]{FBDDF5} 
Togo &
  0.297 \small{\textcolor{teal}{(+0.073)}} &
  \textbf{0.31} \small{\textcolor{purple}{(-0.049)}} &
  0.283 \small{\textcolor{purple}{(-0.173)}} &
  0.252 \small{\textcolor{purple}{(-0.184)}} \\
\rowcolor[HTML]{F1B7EB} 
Bangladesh &
  0.271 \small{\textcolor{teal}{(+0.047)}} &
  \textbf{0.319} \small{\textcolor{purple}{(-0.04)}} &
  0.266 \small{\textcolor{purple}{(-0.19)}} &
  0.221 \small{\textcolor{purple}{(-0.215)}} \\
\rowcolor[HTML]{F1B7EB} 
Bolivia &
  0.3 \small{\textcolor{teal}{(+0.076)}} &
  \textbf{0.318} \small{\textcolor{purple}{(-0.041)}} &
  0.262 \small{\textcolor{purple}{(-0.194)}} &
  0.223 \small{\textcolor{purple}{(-0.213)}} \\
\rowcolor[HTML]{F1B7EB} 
Cambodia &
  \textbf{0.289} \small{\textcolor{teal}{(+0.065)}} &
  0.288 \small{\textcolor{purple}{(-0.071)}} &
  0.213 \small{\textcolor{purple}{(-0.243)}} &
  0.172 \small{\textcolor{purple}{(-0.264)}} \\
\rowcolor[HTML]{F1B7EB} 
Cameroon &
  \textbf{0.313} \small{\textcolor{teal}{(+0.089)}} &
  0.289 \small{\textcolor{purple}{(-0.07)}} &
  0.233 \small{\textcolor{purple}{(-0.223)}} &
  0.192 \small{\textcolor{purple}{(-0.244)}} \\
\rowcolor[HTML]{F1B7EB} 
Cote d'Ivoire &
  0.23 \small{\textcolor{teal}{(+0.006)}} &
  0.296 \small{\textcolor{purple}{(-0.063)}} &
  \textbf{0.325} \small{\textcolor{purple}{(-0.131)}} &
  0.302 \small{\textcolor{purple}{(-0.134)}} \\
\rowcolor[HTML]{F1B7EB} 
Egypt &
  0.257 \small{\textcolor{teal}{(+0.033)}} &
  0.334 \small{\textcolor{purple}{(-0.025)}} &
  \textbf{0.357} \small{\textcolor{purple}{(-0.099)}} &
  0.316 \small{\textcolor{purple}{(-0.12)}} \\
\rowcolor[HTML]{F1B7EB} 
Ghana &
  \textbf{0.314} \small{\textcolor{teal}{ (+0.09)}} &
  0.294 \small{\textcolor{purple}{(-0.065)}} &
  0.267 \small{\textcolor{purple}{(-0.189)}} &
  0.233 \small{\textcolor{purple}{(-0.203)}} \\
\rowcolor[HTML]{F1B7EB} 
Haiti &
  0.296 \small{\textcolor{teal}{(+0.072)}} &
  \textbf{0.331} \small{\textcolor{purple}{(-0.028)}} &
  0.307 \small{\textcolor{purple}{(-0.149)}} &
  0.269 \small{\textcolor{purple}{(-0.167)}} \\
\rowcolor[HTML]{F1B7EB} 
India &
  0.239 \small{\textcolor{teal}{(+0.015)}} &
  \textbf{0.31} \small{\textcolor{purple}{(-0.049)}} &
  0.306 \small{\textcolor{purple}{(-0.15)}} &
  0.278 \small{\textcolor{purple}{(-0.158)}} \\
\rowcolor[HTML]{F1B7EB} 
Iran &
  0.221 \small{\textcolor{purple}{(-0.003)}} &
  0.343 \small{\textcolor{purple}{(-0.016)}} &
  \textbf{0.375} \small{\textcolor{purple}{(-0.081)}} &
  0.337 \small{\textcolor{purple}{(-0.099)}} \\
\rowcolor[HTML]{F1B7EB} 
Jordan &
  0.222 \small{\textcolor{purple}{(-0.002)}} &
  0.308 \small{\textcolor{purple}{(-0.051)}} &
  \textbf{0.376} \small{\textcolor{purple}{(-0.08)}} &
  0.371 \small{\textcolor{purple}{(-0.065)}} \\
\rowcolor[HTML]{F1B7EB} 
Kenya &
  0.296 \small{\textcolor{teal}{(+0.072)}} &
  \textbf{0.318} \small{\textcolor{purple}{(-0.041) }}&
  0.283 \small{\textcolor{purple}{(-0.173)}} &
  0.236 \small{\textcolor{purple}{(-0.2)}} \\
\rowcolor[HTML]{F1B7EB} 
Kyrgyzstan &
  0.229 \small{\textcolor{teal}{(+0.005)}} &
  0.338 \small{\textcolor{purple}{(-0.021)}} &
  \textbf{0.365} \small{\textcolor{purple}{(-0.091)}} &
  0.318 \small{\textcolor{purple}{(-0.118)}} \\
\rowcolor[HTML]{F1B7EB} 
Lebanon &
  0.249 \small{\textcolor{teal}{(+0.025)}} &
  0.309 \small{\textcolor{purple}{(-0.05)}} &
  \textbf{0.348} \small{\textcolor{purple}{(-0.108)}} &
  0.33 \small{\textcolor{purple}{(-0.106)}} \\
\rowcolor[HTML]{F1B7EB} 
Mongolia &
  0.259 \small{\textcolor{teal}{(+0.035)}} &
  \textbf{0.326} \small{\textcolor{purple}{(-0.033)}} &
  0.308 \small{\textcolor{purple}{(-0.148)}} &
  0.256 \small{\textcolor{purple}{(-0.18)}} \\
\rowcolor[HTML]{F1B7EB} 
Myanmar &
  \multicolumn{1}{r}{\cellcolor[HTML]{F1B7EB}0.263 \small{\textcolor{teal}{(+0.039)}}} &
  \multicolumn{1}{r}{\cellcolor[HTML]{F1B7EB}\textbf{0.304} \small{\textcolor{purple}{(-0.055)}}} &
  \multicolumn{1}{r}{\cellcolor[HTML]{F1B7EB}0.241 \small{\textcolor{purple}{(-0.215)}}} &
  \multicolumn{1}{r}{\cellcolor[HTML]{F1B7EB}0.195 \small{\textcolor{purple}{(-0.241)}}} \\
\rowcolor[HTML]{F1B7EB} 
Nepal &
  \multicolumn{1}{r}{\cellcolor[HTML]{F1B7EB}0.274 \small{\textcolor{teal}{(+0.05)}}} &
  \multicolumn{1}{r}{\cellcolor[HTML]{F1B7EB}\textbf{0.307} \small{\textcolor{purple}{(-0.052)}}} &
  \multicolumn{1}{r}{\cellcolor[HTML]{F1B7EB}0.253 \small{\textcolor{purple}{(-0.203)}}} &
  \multicolumn{1}{r}{\cellcolor[HTML]{F1B7EB}0.213 \small{\textcolor{purple}{(-0.223)}}} \\
\rowcolor[HTML]{F1B7EB} 
Nigeria &
  \multicolumn{1}{r}{\cellcolor[HTML]{F1B7EB}\textbf{0.294} \small{\textcolor{teal}{ (+0.07)}}} &
  \multicolumn{1}{r}{\cellcolor[HTML]{F1B7EB}0.286 \small{\textcolor{purple}{(-0.073)}}} &
  \multicolumn{1}{r}{\cellcolor[HTML]{F1B7EB}0.256 \small{\textcolor{purple}{(-0.2)}}} &
  \multicolumn{1}{r}{\cellcolor[HTML]{F1B7EB}0.223 \small{\textcolor{purple}{(-0.213)}}} \\
\rowcolor[HTML]{F1B7EB} 
Pakistan &
  \multicolumn{1}{r}{\cellcolor[HTML]{F1B7EB}0.197 \small{\textcolor{purple}{(-0.027)}}} &
  \multicolumn{1}{r}{\cellcolor[HTML]{F1B7EB}0.303 \small{\textcolor{purple}{(-0.056)}}} &
  \multicolumn{1}{r}{\cellcolor[HTML]{F1B7EB}\textbf{0.321} \small{\textcolor{purple}{(-0.135)}}} &
  \multicolumn{1}{r}{\cellcolor[HTML]{F1B7EB}0.289 \small{\textcolor{purple}{(-0.147)}}} \\
\rowcolor[HTML]{F1B7EB} 
Palestine &
  \multicolumn{1}{r}{\cellcolor[HTML]{F1B7EB}0.258 \small{\textcolor{teal}{(+0.034)}}} &
  \multicolumn{1}{r}{\cellcolor[HTML]{F1B7EB}0.349 \small{\textcolor{purple}{(-0.01)}}} &
  \multicolumn{1}{r}{\cellcolor[HTML]{F1B7EB}\textbf{0.361} \small{\textcolor{purple}{(-0.095)}}} &
  \multicolumn{1}{r}{\cellcolor[HTML]{F1B7EB}0.317 \small{\textcolor{purple}{(-0.119)}}} \\
\rowcolor[HTML]{F1B7EB} 
Papua New Guinea &
  \multicolumn{1}{r}{\cellcolor[HTML]{F1B7EB}0.274 \small{\textcolor{teal}{(+0.05)}}} &
  \multicolumn{1}{r}{\cellcolor[HTML]{F1B7EB}\textbf{0.302} \small{\textcolor{purple}{(-0.057)}}} &
  \multicolumn{1}{r}{\cellcolor[HTML]{F1B7EB}0.266 \small{\textcolor{purple}{(-0.19)}}} &
  \multicolumn{1}{r}{\cellcolor[HTML]{F1B7EB}0.235 \small{\textcolor{purple}{(-0.201)}}} \\
\rowcolor[HTML]{F1B7EB} 
Philippines &
  \multicolumn{1}{r}{\cellcolor[HTML]{F1B7EB}0.271 \small{\textcolor{teal}{(+0.047)}}} &
  \multicolumn{1}{r}{\cellcolor[HTML]{F1B7EB}\textbf{0.346} \small{\textcolor{purple}{(-0.013)}}} &
  \multicolumn{1}{r}{\cellcolor[HTML]{F1B7EB}0.337 \small{\textcolor{purple}{(-0.119)}}} &
  \multicolumn{1}{r}{\cellcolor[HTML]{F1B7EB}0.295 \small{\textcolor{purple}{(-0.141)}}} \\
\rowcolor[HTML]{F1B7EB} 
Sri Lanka &
  \multicolumn{1}{r}{\cellcolor[HTML]{F1B7EB}0.275 \small{\textcolor{teal}{(+0.051)}}} &
  \multicolumn{1}{r}{\cellcolor[HTML]{F1B7EB}\textbf{0.322} \small{\textcolor{purple}{(-0.037)}}} &
  \multicolumn{1}{r}{\cellcolor[HTML]{F1B7EB}0.303 \small{\textcolor{purple}{(-0.153)}}} &
  \multicolumn{1}{r}{\cellcolor[HTML]{F1B7EB}0.277 \small{\textcolor{purple}{(-0.159)}}} \\
\rowcolor[HTML]{F1B7EB} 
Tanzania &
  \multicolumn{1}{r}{\cellcolor[HTML]{F1B7EB}0.287 \small{\textcolor{teal}{(+0.063)}}} &
  \multicolumn{1}{r}{\cellcolor[HTML]{F1B7EB}\textbf{0.292} \small{\textcolor{purple}{(-0.067)}}} &
  \multicolumn{1}{r}{\cellcolor[HTML]{F1B7EB}0.257 \small{\textcolor{purple}{(-0.199)}}} &
  \multicolumn{1}{r}{\cellcolor[HTML]{F1B7EB}0.228 \small{\textcolor{purple}{(-0.208)}}} \\
\rowcolor[HTML]{F1B7EB} 
Tunisia &
  \multicolumn{1}{r}{\cellcolor[HTML]{F1B7EB}0.276 \small{\textcolor{teal}{(+0.052)}}} &
  \multicolumn{1}{r}{\cellcolor[HTML]{F1B7EB}\textbf{0.321} \small{\textcolor{purple}{(-0.038)}}} &
  \multicolumn{1}{r}{\cellcolor[HTML]{F1B7EB}0.314 \small{\textcolor{purple}{(-0.142)}}} &
  \multicolumn{1}{r}{\cellcolor[HTML]{F1B7EB}0.284 \small{\textcolor{purple}{(-0.152)}}} \\
\rowcolor[HTML]{F1B7EB} 
Ukraine &
  \multicolumn{1}{r}{\cellcolor[HTML]{F1B7EB}0.245 \small{\textcolor{teal}{(+0.021)}}} &
  \multicolumn{1}{r}{\cellcolor[HTML]{F1B7EB}0.355 \small{\textcolor{purple}{(-0.004)}}} &
  \multicolumn{1}{r}{\cellcolor[HTML]{F1B7EB}\textbf{0.372} \small{\textcolor{purple}{(-0.084)}}} &
  \multicolumn{1}{r}{\cellcolor[HTML]{F1B7EB}0.323 \small{\textcolor{purple}{(-0.113)}}} \\
\rowcolor[HTML]{F1B7EB} 
Vietnam &
  \multicolumn{1}{r}{\cellcolor[HTML]{F1B7EB}0.229 \small{\textcolor{teal}{(+0.005)}}} &
  \multicolumn{1}{r}{\cellcolor[HTML]{F1B7EB}0.321 \small{\textcolor{purple}{(-0.038)}}} &
  \multicolumn{1}{r}{\cellcolor[HTML]{F1B7EB}\textbf{0.33} \small{\textcolor{purple}{(-0.126)}}} &
  \multicolumn{1}{r}{\cellcolor[HTML]{F1B7EB}0.294 \small{\textcolor{purple}{(-0.142)}}} \\
\rowcolor[HTML]{F1B7EB} 
Zimbabwe &
  \multicolumn{1}{r}{\cellcolor[HTML]{F1B7EB}\textbf{0.312} \small{\textcolor{teal}{(+0.088)}}} &
  \multicolumn{1}{r}{\cellcolor[HTML]{F1B7EB}0.311 \small{\textcolor{purple}{(-0.048)}}} &
  \multicolumn{1}{r}{\cellcolor[HTML]{F1B7EB}0.285 \small{\textcolor{purple}{(-0.171)}}} &
  \multicolumn{1}{r}{\cellcolor[HTML]{F1B7EB}0.242 \small{\textcolor{purple}{(-0.194)}}}
\end{tabular}%
}
\caption{Table of low-income/poor (in lilac) and lower-middle income (in purple) country suffixes and their effect on
Recall for different income groups. For each country suffix, the highest Recall among income groups is highlighted in
bold. The green and red values show how much increase or reduction that country suffix has on the Recall of data
from an income group compared to default English prompts.}
\label{tab:all_trend_table}
\end{table*}

\begin{table*}[ht]
\begin{tabular}{lrrrr}
\hline
\multicolumn{1}{|l|}{}                                          & \multicolumn{4}{c|}{\textbf{Income levels}}                                                                                                         \\ \cline{2-5} 
\multicolumn{1}{|l|}{\multirow{-2}{*}{\textbf{Country Suffix}}} & \multicolumn{1}{l}{\textbf{Poor \hspace{2pt}  $\Delta$}} & \multicolumn{1}{l}{\textbf{Low-mid \hspace{2pt}  $\Delta$}} & \multicolumn{1}{l}{\textbf{Up-mid \hspace{2pt}  $\Delta$}} & \multicolumn{1}{l|}{\textbf{Rich \hspace{2pt}  $\Delta$}} \\
\rowcolor[HTML]{DAE8FC} 
Brazil                                                          & 0.254 \small{\textcolor{teal}{ (+0.03)}}                     & 0.303 \small{\textcolor{purple}{(-0.056)}}                       & \textbf{0.323} \small{\textcolor{purple}{(-0.133)}}             & 0.303 \small{\textcolor{purple}{(-0.133)}}                     \\
\rowcolor[HTML]{DAE8FC} 
China                                                           & 0.213 \small{\textcolor{purple}{(-0.011)}}                    & 0.34 (-0.019)                        & \textbf{0.369} (-0.087)             & 0.319 (-0.117)                     \\
\rowcolor[HTML]{DAE8FC} 
Colombia                                                        & 0.3 \small{\textcolor{teal}{(+0.076)}}                      & \textbf{0.324} \small{\textcolor{purple}{(-0.035)}}              & 0.275 \small{\textcolor{purple}{(-0.181)}}                      & 0.232 \small{\textcolor{purple}{(-0.204)}}                     \\
\rowcolor[HTML]{DAE8FC} 
Guatemala                                                       & 0.269 \small{\textcolor{teal}{(+0.045)}}                    & \textbf{0.314} \small{\textcolor{purple}{(-0.045)}}              & 0.277 \small{\textcolor{purple}{(-0.179)}}                      & 0.233 \small{\textcolor{purple}{(-0.203)}}                     \\
\rowcolor[HTML]{DAE8FC} 
Indonesia                                                       & 0.266 \small{\textcolor{teal}{(+0.042)}}                    & \textbf{0.328} \small{\textcolor{purple}{(-0.031)}}              & 0.303 \small{\textcolor{purple}{(-0.153)}}                      & 0.266 \small{\textcolor{purple}{(-0.17)}}                      \\
\rowcolor[HTML]{DAE8FC} 
Kazakhstan                                                      & 0.254 \small{\textcolor{teal}{(+0.03)}}                     & \textbf{0.337 \small{\textcolor{purple}{(-0.022)}}}              & \textbf{0.337} \small{\textcolor{purple}{(-0.119)}}             & 0.292 \small{\textcolor{purple}{(-0.144)}}                     \\
\rowcolor[HTML]{DAE8FC} 
Mexico                                                          & 0.251 \small{\textcolor{teal}{(+0.027)}}                    & 0.335 \small{\textcolor{purple}{(-0.024)}}                       & \textbf{0.357} \small{\textcolor{purple}{(-0.099)}}             & 0.312 \small{\textcolor{purple}{(-0.124)}}                     \\
\rowcolor[HTML]{DAE8FC} 
Peru                                                            & 0.261 \small{\textcolor{teal}{(+0.037)}}                    & \textbf{0.319} \small{\textcolor{purple}{(-0.04)}}               & 0.317 \small{\textcolor{purple}{(-0.139)}}                      & 0.287 \small{\textcolor{purple}{(-0.149)}}                     \\
\rowcolor[HTML]{DAE8FC} 
Russia                                                          & 0.212 \small{\textcolor{purple}{(-0.012)}}                    & 0.344 \small{\textcolor{purple}{(-0.015)}}                       & \textbf{0.382} \small{\textcolor{purple}{(-0.074)}}             & 0.335 \small{\textcolor{purple}{(-0.101)}}                     \\
\rowcolor[HTML]{DAE8FC} 
Serbia                                                          & 0.197 \small{\textcolor{purple}{(-0.027)}}                    & 0.313 \small{\textcolor{purple}{(-0.046)}}                       & \textbf{0.378} \small{\textcolor{purple}{(-0.078)}}             & 0.354 \small{\textcolor{purple}{(-0.082)}}                     \\
\rowcolor[HTML]{DAE8FC} 
South Africa                                                    & 0.291 \small{\textcolor{teal}{(+0.067)}}                    & \textbf{0.302} \small{\textcolor{purple}{(-0.057)}}              & \textbf{0.302} \small{\textcolor{purple}{(-0.154)}}             & 0.269 \small{\textcolor{purple}{(-0.167)}}                     \\
\rowcolor[HTML]{DAE8FC} 
Thailand                                                        & 0.234 \small{\textcolor{teal}{(+0.01)}}                     & \textbf{0.312} \small{\textcolor{purple}{(-0.047)}}              & 0.293 \small{\textcolor{purple}{(-0.163)}}                      & 0.256 \small{\textcolor{purple}{(-0.18)}}                      \\
\rowcolor[HTML]{DAE8FC} 
Turkey                                                          & 0.228 \small{\textcolor{teal}{(+0.004)}}                    & 0.321 \small{\textcolor{purple}{(-0.038)}}                       & \textbf{0.333} \small{\textcolor{purple}{(-0.123)}}             & 0.302 \small{\textcolor{purple}{(-0.134)}}                     \\
\rowcolor[HTML]{CBCEFB} 
Austria                                                         & 0.166 \small{\textcolor{purple}{(-0.058)}}                    & 0.296 \small{\textcolor{purple}{(-0.063)}}                       & 0.407 \small{\textcolor{purple}{(-0.049)}}                      & \textbf{0.408} \small{\textcolor{purple}{(-0.028)}}            \\
\rowcolor[HTML]{CBCEFB} 
Canada                                                          & 0.266 \small{\textcolor{teal}{(+0.042)}}                    & 0.355 \small{\textcolor{purple}{(-0.004)}}                       & \textbf{0.391} \small{\textcolor{purple}{(-0.065)}}             & 0.359 \small{\textcolor{purple}{(-0.077)}}                     \\
\rowcolor[HTML]{CBCEFB} 
Czech Republic                                                  & 0.195 \small{\textcolor{purple}{(-0.029)}}                    & 0.33 \small{\textcolor{purple}{(-0.029)}}                        & \textbf{0.395} \small{\textcolor{purple}{(-0.061)}}             & 0.379 \small{\textcolor{purple}{(-0.057)}}                     \\
\rowcolor[HTML]{CBCEFB} 
Denmark                                                         & 0.184 \small{\textcolor{purple}{(-0.04)}}                     & 0.293 \small{\textcolor{purple}{(-0.066)}}                       & 0.386 \small{\textcolor{purple}{(-0.07)}}                       & \textbf{0.394} \small{\textcolor{purple}{(-0.042)}}            \\
\rowcolor[HTML]{CBCEFB} 
France                                                          & 0.199 \small{\textcolor{purple}{(-0.025)}}                    & 0.317 \small{\textcolor{purple}{(-0.042)}}                       & 0.415 \small{\textcolor{purple}{(-0.041)}}                      & \textbf{0.417} \small{\textcolor{purple}{(-0.019)}}            \\
\rowcolor[HTML]{CBCEFB} 
Italy                                                           & 0.192 \small{\textcolor{purple}{(-0.032)}}                    & 0.318 \small{\textcolor{purple}{(-0.041)}}                       & \textbf{0.379} \small{\textcolor{purple}{(-0.077)}}             & 0.367 \small{\textcolor{purple}{(-0.069)}}                     \\
\rowcolor[HTML]{CBCEFB} 
Netherlands                                                     & 0.219 \small{\textcolor{purple}{(-0.005)}}                    & 0.31 \small{\textcolor{purple}{(-0.049)}}                        & \textbf{0.374} \small{\textcolor{purple}{(-0.082)}}             & 0.359 \small{\textcolor{purple}{(-0.077)}}                     \\
\rowcolor[HTML]{CBCEFB} 
Romania                                                         & 0.255 \small{\textcolor{teal}{(+0.031)}}                    & 0.337 \small{\textcolor{purple}{(-0.022)}}                       & \textbf{0.342} \small{\textcolor{purple}{(-0.114)}}             & 0.312 \small{\textcolor{purple}{(-0.124)}}                     \\
\rowcolor[HTML]{CBCEFB} 
South Korea                                                     & 0.225 \small{\textcolor{teal}{(+0.001)}}                    & 0.313 \small{\textcolor{purple}{(-0.046)}}                       & \textbf{0.345} \small{\textcolor{purple}{(-0.111)}}             & 0.314 \small{\textcolor{purple}{(-0.122)}}                     \\
\rowcolor[HTML]{CBCEFB} 
Spain                                                           & 0.183 \small{\textcolor{purple}{(-0.041)}}                    & 0.308 \small{\textcolor{purple}{(-0.051)}}                       & \textbf{0.413} \small{\textcolor{purple}{(-0.043)}}             & 0.404 \small{\textcolor{purple}{(-0.032)}}                     \\
\rowcolor[HTML]{CBCEFB} 
Sweden                                                          & 0.167 \small{\textcolor{purple}{(-0.057)}}                    & 0.294 \small{\textcolor{purple}{(-0.065)}}                       & 0.405 \small{\textcolor{purple}{(-0.051)}}                      & \textbf{0.412} \small{\textcolor{purple}{(-0.024)}}            \\
\rowcolor[HTML]{CBCEFB} 
Switzerland                                                     & 0.135 \small{\textcolor{purple}{(-0.089)}}                    & 0.257 \small{\textcolor{purple}{(-0.102)}}                       & 0.363 \small{\textcolor{purple}{(-0.093)}}                      & \textbf{0.389} \small{\textcolor{purple}{(-0.047)}}            \\
\rowcolor[HTML]{CBCEFB} 
United Kingdom                                                  & 0.205 \small{\textcolor{purple}{(-0.019)}}                    & 0.326 \small{\textcolor{purple}{(-0.033)}}                       & \textbf{0.418} \small{\textcolor{purple}{(-0.038)}}             & 0.409 \small{\textcolor{purple}{(-0.027)}}                     \\
\rowcolor[HTML]{CBCEFB} 
United States                                                   & 0.25 \small{\textcolor{teal}{(+0.026)}}                     & 0.362 \small{\textcolor{purple}{(-0.003)}}                       & \textbf{0.421} \small{\textcolor{purple}{(-0.035)}}             & 0.391 \small{\textcolor{purple}{(0.045)}}                     
\end{tabular}
\caption{Table of high-income/rich (in blue) and upper-middle-income (in sky blue) country suffixes and their effect on Recall for different income groups. For each country suffix, the highest Recall among income groups is highlighted in bold. The green and red values show how much increase or reduction that country suffix has on the Recall of data from an income group compared to default English prompts.}
\label{tab:trend_table_2}
\end{table*}


\begin{figure*}[ht]
    \centering
    \includegraphics[width=\textwidth]{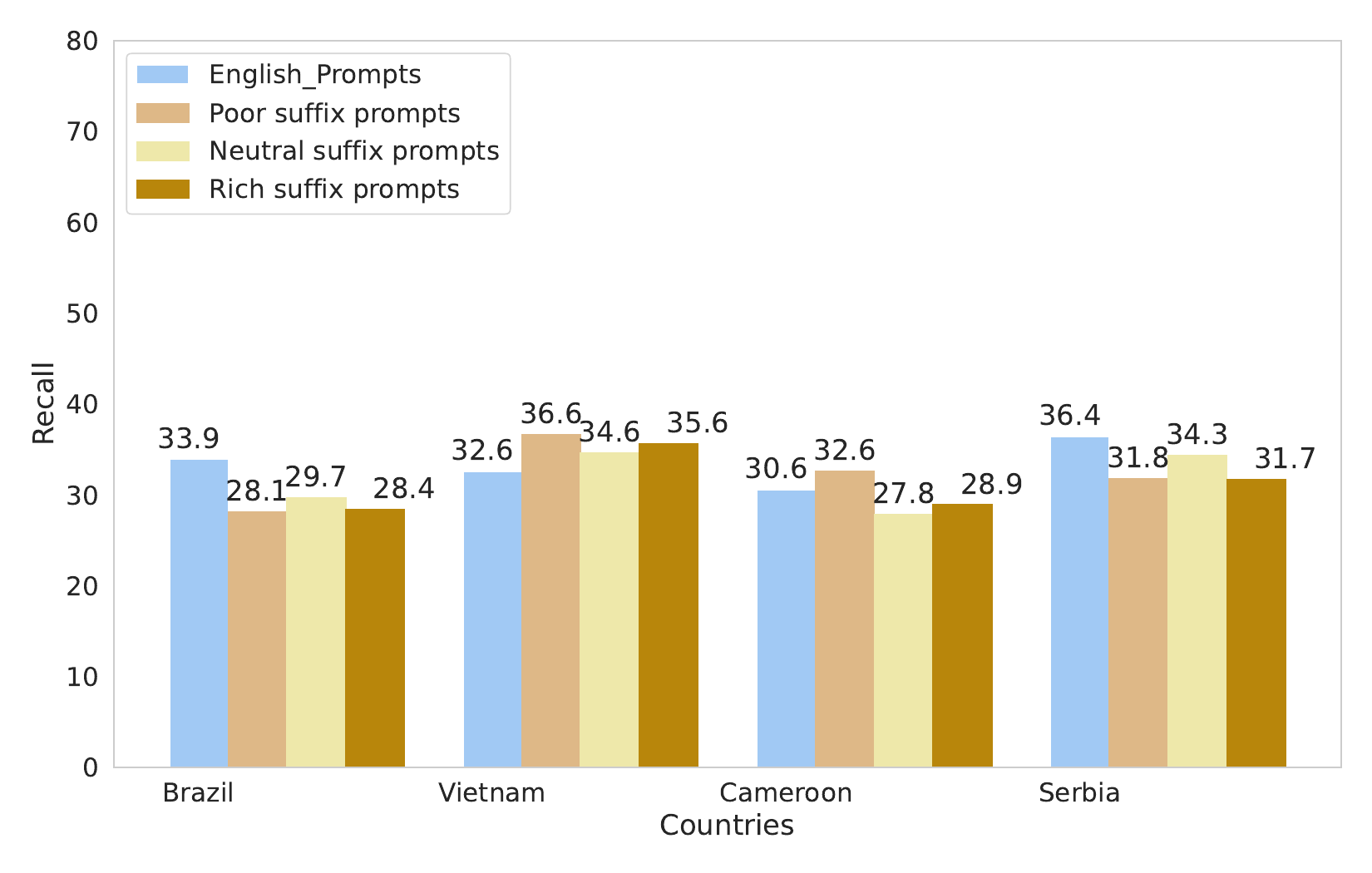}
    \caption{Average NLLB SigLIP Recall over poor and lower-middle income images  for English and Income Suffix prompts}
    \label{fig:NLLBIncomeisbest}
\end{figure*}

\begin{table}[]
\resizebox{\columnwidth}{!}{%
\begin{tabular}{|
>{\columncolor[HTML]{FFFFFF}}l |
>{\columncolor[HTML]{FFFFFF}}l |
>{\columncolor[HTML]{FFFFFF}}r |}
\hline
\textbf{Language} & \textbf{ISO} & \multicolumn{1}{l|}{\cellcolor[HTML]{FFFFFF}\textbf{chrf++}} \\ \hline
Arabic            & arb\_Arab    & 51.4                                                         \\ \hline
Bengali           & ben\_Beng    & 46.2                                                         \\ \hline
Burmese           & mya\_Mymr    & 29.3                                                         \\ \hline
Chinese           & zho\_Hans    & 19.6                                                         \\ \hline
Creole            & hat\_Latn    & 50.2                                                         \\ \hline
Czech             & ces\_Latn    & 52.7                                                         \\ \hline
Danish            & dan\_Latn    & 63.2                                                         \\ \hline
Dutch             & nld\_Latn    & 53.1                                                         \\ \hline
Ewe               & ewe\_Latn    & 35.6                                                         \\ \hline
Farsi\_Persian    & pes\_Arab    & 47.4                                                         \\ \hline
French            & fra\_Latn    & 67.0                                                         \\ \hline
German            & deu\_Latn    & 59.4                                                         \\ \hline
Hausa             & hau\_Latn    & 49.0                                                         \\ \hline
Hindi             & hin\_Latn    & 54.2                                                         \\ \hline
Indonesian        & ind\_Latn    & 66.6                                                         \\ \hline
Italian           & ita\_Latn    & 54.6                                                         \\ \hline
Khmer             & khm\_Khmr    & 31.2                                                         \\ \hline
Kinyarwanda       & kin\_Latn    & 44.0                                                         \\ \hline
Korean            & kor\_Hang    & 32.1                                                         \\ \hline
Kyrgyz            & kir\_Cyrl    & 42.6                                                         \\ \hline
Mongolian         & khk\_Cyrl    & 37.3                                                         \\ \hline
Nepali            & npi\_Deva    & 49.0                                                         \\ \hline
Oromo             & gaz\_Latn    & 31.6                                                         \\ \hline
Portuguese        & por\_Latn    & 67.4                                                         \\ \hline
Romanian          & ron\_Latn    & 58.2                                                         \\ \hline
Russian           & rus\_Cyrl    & 52.5                                                         \\ \hline
Serbian           & srp\_Cyrl    & 53.3                                                         \\ \hline
Shona             & sna\_Latn    & 42.9                                                         \\ \hline
Sinhala           & sin\_Sinh    & 42.4                                                         \\ \hline
Somali            & som\_Latn    & 41.5                                                         \\ \hline
Spanish           & spa\_Latn    & 52.6                                                         \\ \hline
Swahili           & swh\_Latn    & 58.0                                                         \\ \hline
Swedish           & swe\_Latn    & 62.7                                                         \\ \hline
Tagalog           & tgl\_Latn    & 56.4                                                         \\ \hline
Thai              & tha\_Thai    & 36.0                                                         \\ \hline
Turkish           & tur\_Latn    & 52.9                                                         \\ \hline
Ukrainian         & ukr\_Cyrl    & 50.5                                                         \\ \hline
Urdu              & urd\_Arab    & 46.6                                                         \\ \hline
Vietnamese        & vie\_Latn    & 56.4                                                         \\ \hline
Zulu              & zul\_Latn    & 51.0                                                         \\ \hline
\end{tabular}%
}
\caption{Languages used and translation metrics (chrf++ scores) for  NLLB-200-distilled-600M from English to these languages.}
\label{tab:Translation metrics}
\end{table}



\begin{table}[]
\centering
\resizebox{\columnwidth}{!}{%
\begin{tabular}{l|ll}
\hline
\textbf{Prompt}                  & \textbf{P-value} & \textbf{Sig. or not} \\ \hline
English \& Native translated     & 8.64e-09         & yes                  \\
English \& Country suffix        & 7.27e-08         & yes                  \\
English \& Poor Income Suffix    & 0.02             & yes                  \\
English \& Rich Income Suffix    & 0.603            & no                   \\
English \& Neutral Income Suffix & 0.563            & no                  
\end{tabular}%
}
\caption{Table showing p-values of Wilcoxon test between the default English prompt and each of the formulated prompts. The difference is regarded as statistically significant when p $\le$ 0.05.}
\label{tab:wilcoxon}
\end{table}

\end{document}